\documentclass[aps,prd,showpacs,twocolumn,superscriptaddress]{revtex4-1}
\usepackage{graphicx}
\usepackage{epsfig,graphics,subfigure,psfrag,amsmath,amssymb}
\usepackage{lineno}
\usepackage{dcolumn}
\usepackage{bm}
\usepackage{overpic}
\usepackage{xspace}
\usepackage{rotating}
\usepackage{color}
\usepackage{multirow,diagbox}
\usepackage{colortbl}
\usepackage{epstopdf}
\usepackage{lipsum}
\usepackage{mathtools}
\usepackage{cuted}
\usepackage[utf8]{inputenc}
\usepackage[colorlinks,linkcolor=blue,anchorcolor=blue,citecolor=blue,urlcolor=blue]{hyperref}
\usepackage[table]{xcolor}

\newcommand{\eff}{\varepsilon}
\newcommand{\jpsi}{J/\psi}
\newcommand{\pip}{\pi^{+}}
\newcommand{\pim}{\pi^{-}}
\newcommand{\etap}{\eta^{\prime}}
\newcommand{\chisq}{\chi^{2}}

\newcommand{\ksks}{K_{S}^{0}K_{S}^{0}}

\newcommand{\gam}{\gamma}

\newcommand{\pp}{\pi^{+}\pi^{-}}

\newcommand{\piz}{\pi^{0}}

\newcommand{\bietap}{\etap\etap}
\newcommand{\doube}{\eta\eta}

\begin{document}
\title{\boldmath Partial wave analysis of $J/\psi \to \gamma \eta^{\prime} \eta^{\prime}$ }

\author{
M.~Ablikim$^{1}$, M.~N.~Achasov$^{10,b}$, P.~Adlarson$^{68}$, S. ~Ahmed$^{14}$, M.~Albrecht$^{4}$, R.~Aliberti$^{28}$, A.~Amoroso$^{67A,67C}$, M.~R.~An$^{32}$, Q.~An$^{64,50}$, X.~H.~Bai$^{58}$
% ...
\\
\vspace{0.2cm}
(BESIII Collaboration)\\
\vspace{0.2cm} {\it
$^{1}$ Institute of High Energy Physics, Beijing 100049, People's Republic of China\\
% ...
$^{4}$ Bochum Ruhr-University, D-44780 Bochum, Germany\\
% ...
$^{10}$ G.I. Budker Institute of Nuclear Physics SB RAS (BINP), Novosibirsk 630090, Russia\\
% ...
$^{14}$ Helmholtz Institute Mainz, Staudinger Weg 18, D-55099 Mainz, Germany\\
% ...
$^{28}$ Johannes Gutenberg University of Mainz, Johann-Joachim-Becher-Weg 45, D-55099 Mainz, Germany\\
% ...
$^{32}$ Liaoning Normal University, Dalian 116029, People's Republic of China\\
% ...
$^{50}$ State Key Laboratory of Particle Detection and Electronics, Beijing 100049, Hefei 230026, People's Republic of China\\
% ...
$^{58}$ University of Jinan, Jinan 250022, People's Republic of China\\
% ...
$^{64}$ University of Science and Technology of China, Hefei 230026, People's Republic of China\\
% ...
$^{67}$ University of Turin and INFN, (A)University of Turin, I-10125, Turin, Italy; (B)University of Eastern Piedmont, I-15121, Alessandria, Italy; (C)INFN, I-10125, Turin, Italy\\
$^{68}$ Uppsala University, Box 516, SE-75120 Uppsala, Sweden\\
% ...
$^{b}$ Also at the Novosibirsk State University, Novosibirsk, 630090, Russia\\
% ...
}
}
\date{\today}

\begin{abstract}
Using a sample of $(10.09~\pm~0.04)\times10^{9} ~J/\psi$ events collected with the BESIII detector, a partial wave analysis of $J/\psi\to\gamma\etap\etap$ is performed. 
The masses and widths of the observed resonances and their branching fractions are reported.
The main contribution is from $\jpsi\rightarrow\gam f_0(2020)$ with $f_0(2020)\rightarrow\etap\etap$, which is found with a significance of greater than 25$\sigma$. 
The product branching fraction ${\cal B}\left(\jpsi\rightarrow\gam f_0(2020)\right)\cdot{\cal B}\left(f_0(2020)\rightarrow\etap\etap\right)$ is
measured to be $(2.63\pm0.06({\rm stat.})^{+0.31}_{-0.46}({\rm syst.}))\times10^{-4}$.
\end{abstract}

\pacs{13.25.Gv, 14.40.Be}
\maketitle

\section{Introduction}
Due to the non-Abelian structure of quantum chromodynamics (QCD), bound states beyond those in the constituent quark model, such as glueballs, which are formed by gluons, are expected~\cite{Amsler:2004ps,Klempt:2007cp,Crede:2008vw}.
The identification of glueballs would provide validation for the quantitative understanding of QCD and the study of glueballs thus plays an important role in the field of hadron physics.
However, possible mixing of pure glueballs with nearby $q\bar{q}$ nonet mesons makes the identification of glueballs difficult both experimentally and theoretically.
The radiative $\jpsi$ decay is a gluon-rich process and is therefore regarded as one of the most promising hunting grounds for glueballs~\cite{Sarantsev:2021ein,Rodas:2021tyb}.
Searching for glueballs in $\jpsi\rightarrow\gam\etap\etap$, a decay mode which has not been previously explored, is essential.

The spectrum of glueballs is predicted by quenched Lattice QCD (LQCD) with the lightest candidate having scalar quantum numbers $0^{++}$ and a mass in the range of 1.5$-$1.7~GeV/$c^{2}$ with its first excitation at a mass of around 2.6~GeV/$c^{2}$~\cite{Bali:1993fb,Morningstar:1999rf,Chen:2005mg,Gregory:2012hu,Sun:2017ipk}.
In the partial wave analyses (PWA) of $\jpsi\rightarrow\gam\eta\eta$~\cite{Ablikim:2013hq} and $\jpsi\rightarrow\gam\ksks$~\cite{Ablikim:2018izx}, the production rate of $f_0(1710)$ is one order of magnitude larger than that of $f_0(1500)$.
Referring to the measurements of radiative decays of $\jpsi$ to two mesons
listed by the Particle Data Group (PDG)~\cite{Zyla:2020zbs}, 
the expected decay rate of $\jpsi\rightarrow\gam f_0(1710)$ is larger than $2.1\times 10^{-3}$, which is comparable to the LQCD prediction of a scalar glueball 
with rate of $3.8\times 10^{-3}$~\cite{Gui:2012gx}.

The high production rate suggests that the $f_0(1710)$ has a large gluonic component.
The copious production of scalar resonances around 2.1~GeV/$c^2$ has also been reported in $\jpsi\rightarrow\gam\doube$~\cite{Ablikim:2013hq} and $\jpsi\rightarrow\gam\ksks$~\cite{Ablikim:2018izx}.
The high production rate of these scalar resonances, which is similar to that of $f_0(1710)$, indicates that the scalars around 2.1~GeV/$c^2$ may significantly overlap with the first scalar glueball excitation.
The decay width of a scalar glueball to $\etap\etap$ should be comparable to that of a scalar glueball to $\eta\eta$, considering flavor symmetry.
Studies from $p\bar{p}$ annihilation, another gluon-rich process, show that the $f_0(2100)$ has a coupling to $\eta\eta$ stronger than that to $\pi\pi$
~\cite{Anisovich:2000ut,Anisovich:2000ae}, revealing a strange decay pattern.
Searching for scalars around 2.1 GeV/$c^2$ in $\etap\etap$ will help clarify their nature.

The mass of the lightest tensor glueball state is predicted to be 2.3$-$2.4~GeV/$c^{2}$~\cite{Morningstar:1999rf,Chen:2005mg}, which is consistent with the measurement of the $f_2(2340)$.
The $f_2(2340)$ was observed
firstly
in $\pim p \rightarrow\phi\phi n$~\cite{Etkin:1987rj}. 
It was also observed in $\jpsi\rightarrow\gam\eta\eta$~\cite{Ablikim:2013hq} and $\jpsi\rightarrow\gam\phi\phi$~\cite{Ablikim:2016hlu} with large production rates.
A significant tensor structure around 2.4~GeV/$c^2$ was also seen in $\jpsi\rightarrow\gam\piz\piz$~\cite{Ablikim:2015umt} and $\jpsi\rightarrow\gam\ksks$~\cite{Ablikim:2018izx}.
However, the measured production rate of the $f_2(2340)$ in radiative $\jpsi$ decays from existing experimental studies is much lower than that from the
LQCD predictions of the tensor glueball~\cite{Yang:2013xba}.
Studies of more decay modes of the $f_2(2340)$ are desirable.

In this paper, the result of a partial wave analysis of $\jpsi\rightarrow\gam\etap\etap$ is presented based on a sample of $10.09\times10^9~\jpsi$ events~\cite{BESIII:2021cxx} collected with the BESIII detector.
The two $\etap$ are reconstructed using their decays to $\gam\pp$ or $\eta(\rightarrow\gam\gam)\pp$.
Two combinations of $\etap$ decays are used to reconstruct $\jpsi\rightarrow\gam\bietap$:
in the first, called mode~I, both $\etap$ are from $\eta\pp$; 
in the second, mode~II, one $\etap$ decays to $\gam\pp$ and the other $\etap$ decays to $\eta\pp$.

\section{Detector and Monte Carlo simulations}
The BESIII detector~\cite{Ablikim:2009aa} records symmetric $e^+e^-$ collisions provided by the BEPCII storage ring 
with a designed peak 
luminosity of $1\times10^{33}$~cm$^{-2}$s$^{-1}$ in the center-of-mass energy range from 2.0 to 4.9~GeV.
BESIII has collected large data samples in this energy region~\cite{BESIII:2020nme}. 
The cylindrical core of the BESIII detector covers 93\% of the full solid angle and consists of a helium-based multilayer drift chamber~(MDC), a plastic scintillator time-of-flight system~(TOF), and a CsI(Tl) electromagnetic calorimeter~(EMC), which are all enclosed in a superconducting solenoidal magnet providing a 1.0~T (0.9~T in 2012) magnetic field. 
The solenoid is supported by an octagonal flux-return yoke with resistive plate counter muon identification modules interleaved with steel.
The charged-particle momentum resolution at $1~{\rm GeV}/c$ is $0.5\%$, and the d$E$/d$x$ resolution is $6\%$ for electrons from Bhabha scattering. 
The EMC measures photon energies with a resolution of $2.5\%$ ($5\%$) at $1$~GeV in the barrel (end cap) region. 
The time resolution in the TOF barrel region is 68~ps, while that in the end cap region was initially 110~ps.
The end cap TOF system was upgraded in 2015 with multi-gap resistive plate chamber technology, improving the time resolution to be 60 ps~\cite{Li:2017,*Guo:2017,*Cao:2020ibk}.

Simulated data samples produced with a {\sc geant4}-based~\cite{Agostinelli:2002hh} Monte Carlo (MC) package, which includes the geometric description of the BESIII detector and the detector response, are used to determine detection efficiencies and to estimate backgrounds. 
The simulation models the beam energy spread and initial state radiation (ISR) in the $e^+e^-$ annihilations with the generator {\sc kkmc}~\cite{Jadach:2000ir,*Jadach:1999vf}.
The inclusive MC sample includes both the production of the $J/\psi$ resonance and the continuum processes incorporated in {\sc kkmc}~\cite{Jadach:2000ir,*Jadach:1999vf}.
The known decay modes are modeled with {\sc evtgen}~\cite{Lange:2001uf,*Ping:2008zz} using branching fractions taken from the PDG~\cite{Zyla:2020zbs}, and the remaining unknown charmonium decays are modeled with {\sc lundcharm}~\cite{Chen:2000tv,*Yang:2014vra}. 
Final state radiation (FSR) from charged final state particles is incorporated using the {\sc photos} package~\cite{RichterWas:1992qb}.

To estimate the detection efficiency and to optimize the selection criteria, signal MC events are generated for $\jpsi\rightarrow\gam\etap\etap$ in mode I and II.
The process $\etap\to\gamma\pi^{+}\pi^{-}$ is simulated taking into account both the $\rho - \omega$ interference and the box anomaly~\cite{Ablikim:2017fll}.
The remaining processes excluding $\etap\rightarrow\gam\pip\pim$ are simulated using a phase-space (PHSP) generator.
Based on studies of Bhabha, di-muon and inclusive hadronic event samples, the trigger efficiency for events with charged tracks is found to be approximately 100\%~\cite{Ablikim:2020vku,Berger:2010my} and is thus neglected in this analysis.

\section{Event selection}
Charged tracks detected in the MDC are required to be within a polar angle ($\theta$) range of $|\rm{cos~\theta}|<0.93$, where $\theta$ is defined with respect to the $z$-axis.
The distance of closest approach to the interaction point (IP) must be less than 10\,cm along the $z$-axis, $|V_{z}|$, and less than 1\,cm in the transverse plane, $|V_{xy}|$.
Particle identification~(PID) for charged tracks combines measurements of the energy deposited in the MDC~(d$E$/d$x$) and the flight time in the TOF to form likelihoods $\mathcal{L}(h)~(h=p,K,\pi)$ for each hadron $h$ hypothesis.
Each track is assigned to the particle type corresponding to the highest likelihoods.
Candidate events are required to have four charged $\pi$ tracks and zero net charge.
Photon candidates are identified using showers in the EMC. 
The deposited energy of each shower must be more than 25~MeV in the barrel region ($|\cos \theta|< 0.80$) and more than 50~MeV in the end cap region ($0.86 <|\cos \theta|< 0.92$).
To exclude showers that originate from charged tracks, the angle between the position of each shower in the EMC and the closest extrapolated charged track must be greater than $10^{\circ}$.
To suppress electronic noise and showers unrelated to the event, the difference between the EMC time and the event start time is required to be within (0, 700) ns.

\begin{figure}[htbp!]
	\centering
	\includegraphics[width=0.25\textwidth]{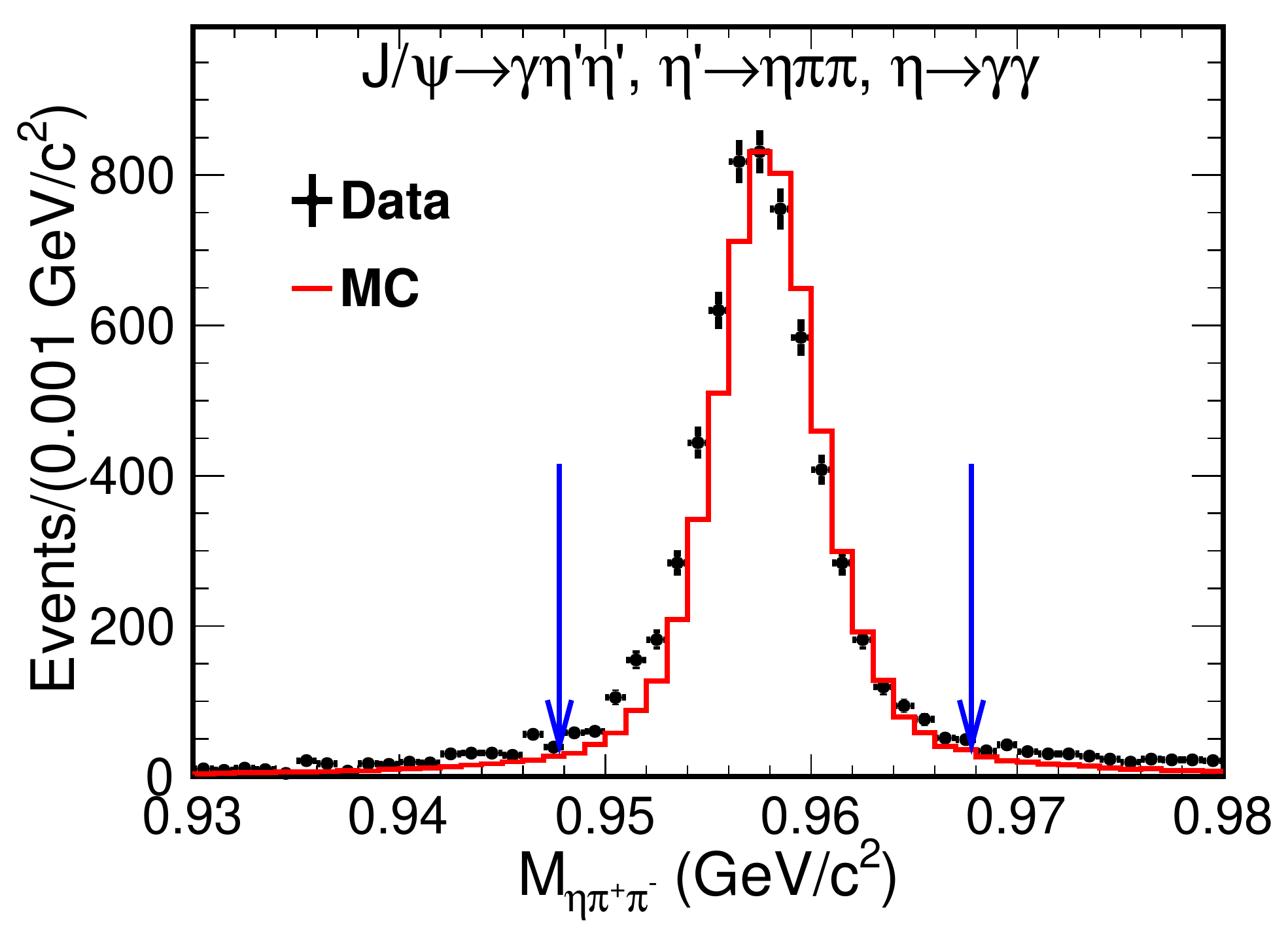} \put(-22,74){(a)}\\
	\includegraphics[width=0.25\textwidth]{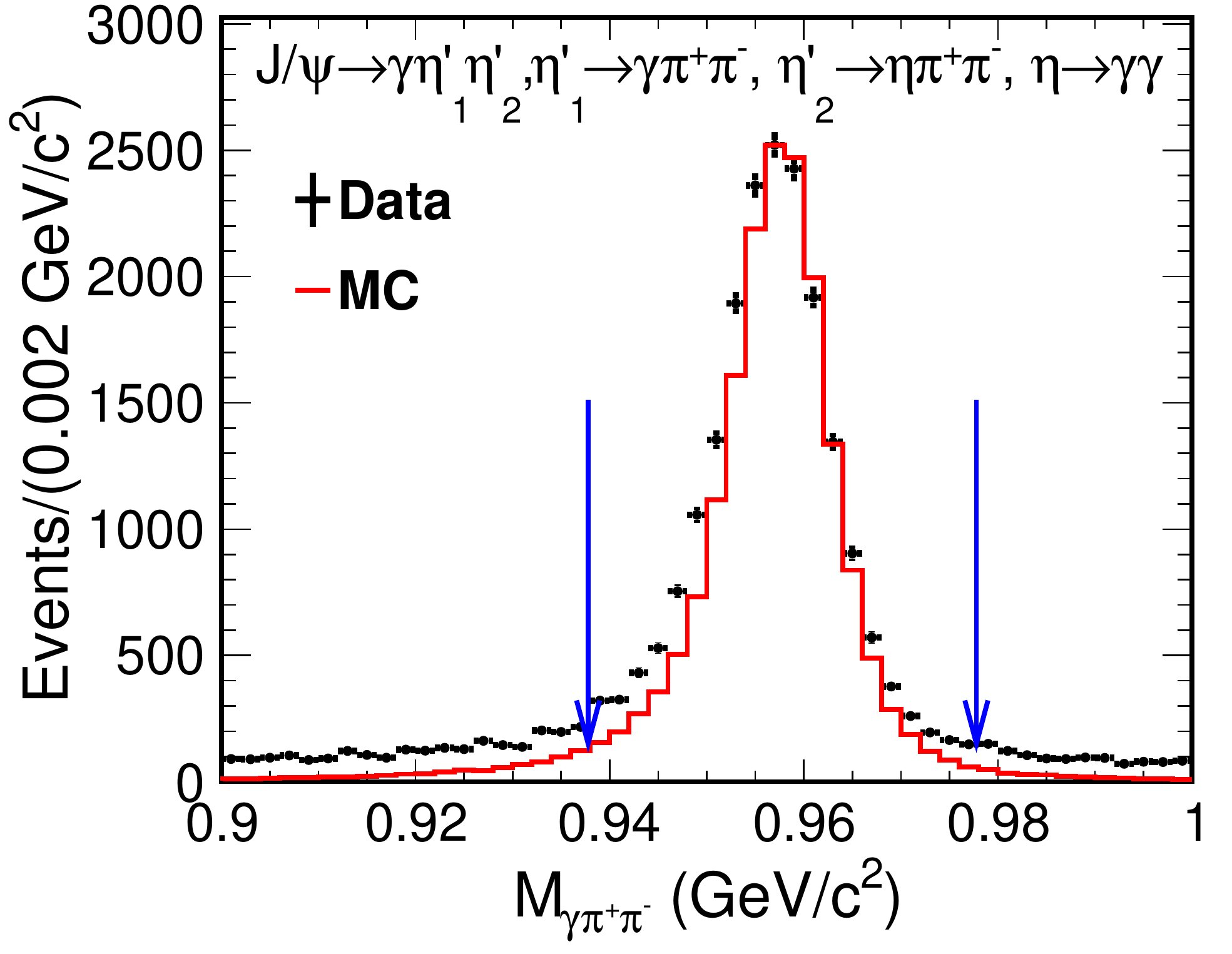} \put(-22,74){(b)}
	\includegraphics[width=0.25\textwidth]{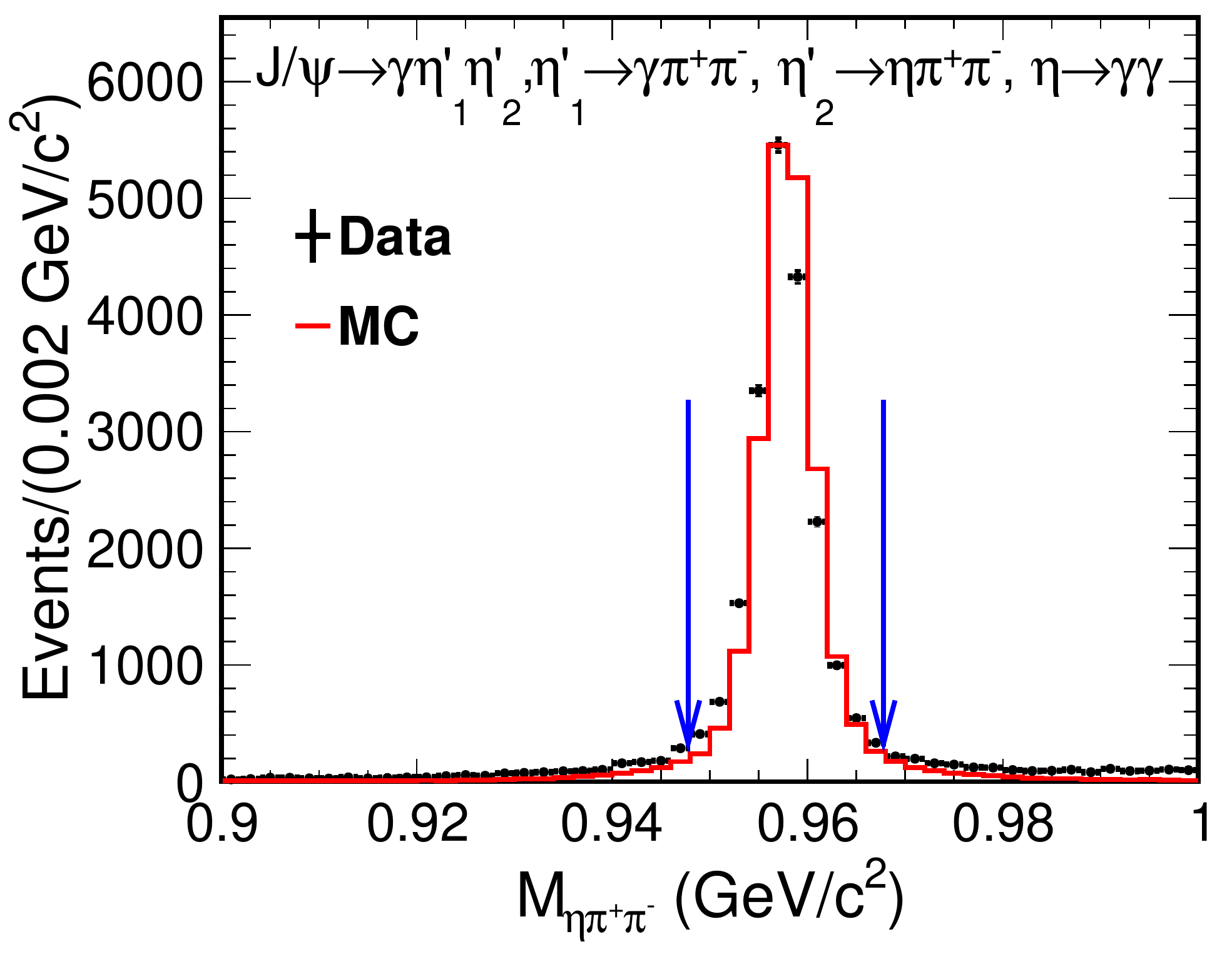} \put(-20,74){(c)}
	\caption{
	\label{fig:metap}
	Invariant mass distributions of (a) $M_{\eta\pp}$ in mode I, and (b) $M_{\gam\pp}$, (c) $M_{\eta\pp}$ in mode II for the selected candidates.
	The dots with error bars are data and the histograms depict the signal MC samples (normalized by height).}
\end{figure}

\begin{figure*}[htbp!]
	\centering
	\includegraphics[width=0.45\textwidth]{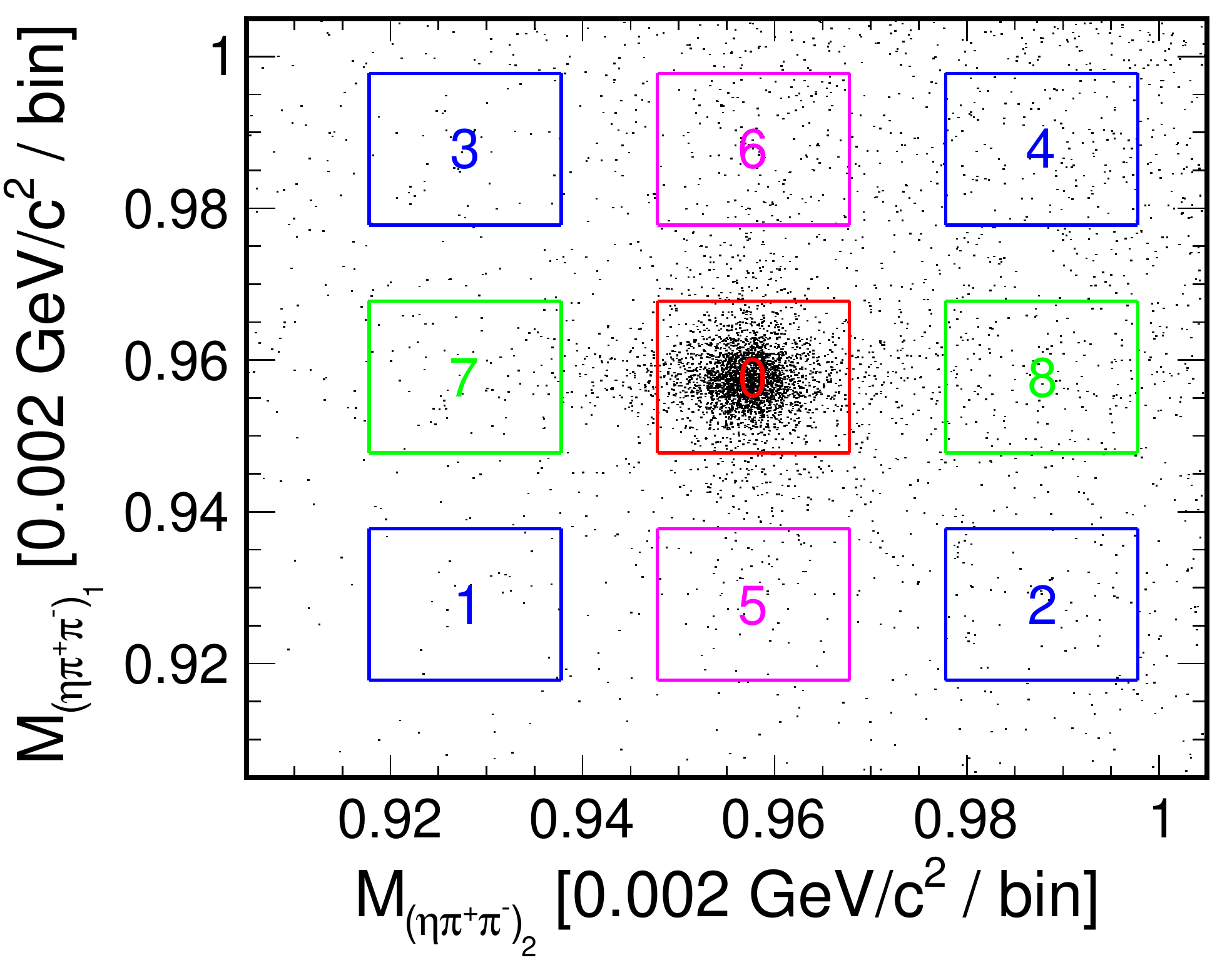} \put(-113,-8){(a)}
	\hspace*{2mm}
	\includegraphics[width=0.46\textwidth]{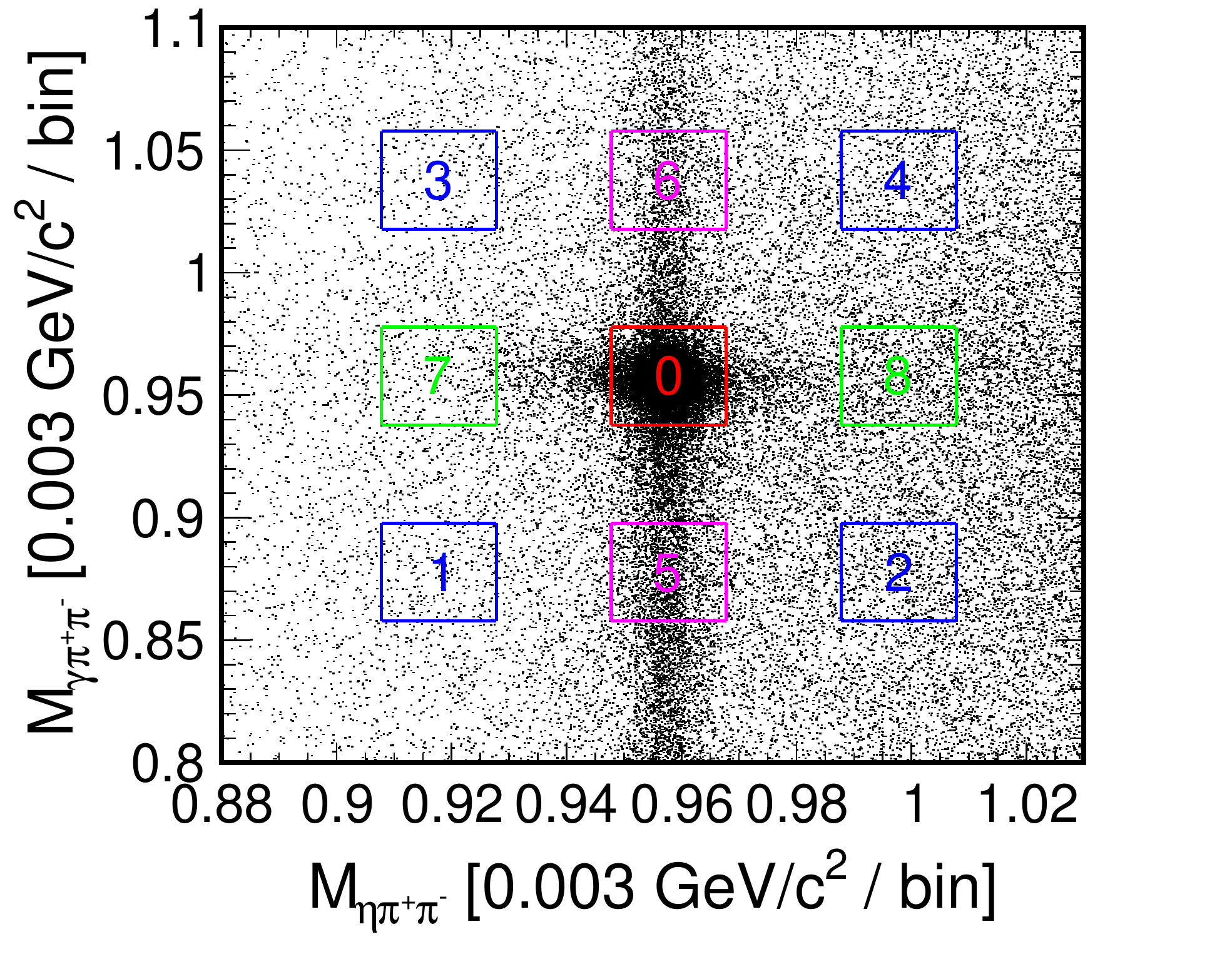} \put(-105,-8){(b)}
	\caption{\label{fig:sdb_divide}
	The distributions of (a) $M_{(\eta\pp)_1}$ versus $M_{(\eta\pp)_2}$ in mode I and (b) $M_{\gam\pp}$ versus $M_{\eta\pp}$ in mode II. 
	The red box labeled as ``0'' and other boxes labeled from ``1'' to ``8'' are signal region and sideband regions as defined in the text, respectively.}
\end{figure*}

In mode I, with $\jpsi\rightarrow\gam\etap\etap$, $\etap\rightarrow\eta\pp$, and $\eta\rightarrow\gam\gam$, the final state consists of 5$\gam$ and 2($\pp$).
To reduce background events and improve mass resolution, a six-constraint (6C) kinematic fit is performed under the hypothesis of $\jpsi\rightarrow\gam\eta\eta\pp\pp$ imposing energy-momentum conservation (4C) and constraining the mass of each pair of photons to the nominal mass of $\eta$~\cite{Zyla:2020zbs}.
For events with more than one combination of $\jpsi\rightarrow\gam\eta\eta\pp\pp$, the combination with the least $\chi_{\rm 6C}^{2}$ is selected, and $\chi_{\rm 6C}^{2}<$ 85 is required.
To suppress background contributions with one more or one less photon, four-constraint kinematic fits are performed separately under the hypotheses of $\jpsi\rightarrow4\gam\pp\pp,~\jpsi\rightarrow5\gam\pp\pp,~\jpsi\rightarrow6\gam\pp\pp$, whose $\chisq$ values are denoted as $\chisq_{4\gam},~\chisq_{5\gam},~\chisq_{6\gam}$, respectively.
Events with $\chisq_{5\gam}<\chisq_{4\gam}$ and $\chisq_{5\gam}<\chisq_{6\gam}$ are accepted.
Two $\etap$ candidates are reconstructed from the $\eta\pp$ combinations with the least $\left|M_{(\eta\pp)_1}-m_{\etap}\right|^2+\left|M_{(\eta\pp)_2}-m_{\etap}\right|^2$,
where $M_{(\eta\pp)_{1,2}}$ are the invariant masses of different $\eta\pp$ combinations and $m_{\etap}$ is the nominal mass of the $\etap$~\cite{Zyla:2020zbs}.
The $\etap$ candidates are then selected with $|M_{\eta\pi^{+}\pi^{-}}-m_{\etap}|<$ 0.01~GeV/$c^{2}$.
Here, the two $\etap$ candidates are denoted as $\etap_1$ and $\etap_2$, based on the absolute value of their momenta, for which $\left|\vec{p}_{\etap_1}\right| > \left|\vec{p}_{\etap_2}\right|$.
A clear $\etap$ signal is observed in the invariant mass distribution of $\eta\pp$ ($M_{\eta\pp}$), as shown in Fig.~\ref{fig:metap}(a).
The distribution of the two $M_{\eta\pp}$ of the remaining events is shown in Fig.~\ref{fig:sdb_divide}(a), where the signal region is the red box labeled as ``0''.

\begin{figure*}[htbp!]
	\centering
	\includegraphics[width=0.45\textwidth]{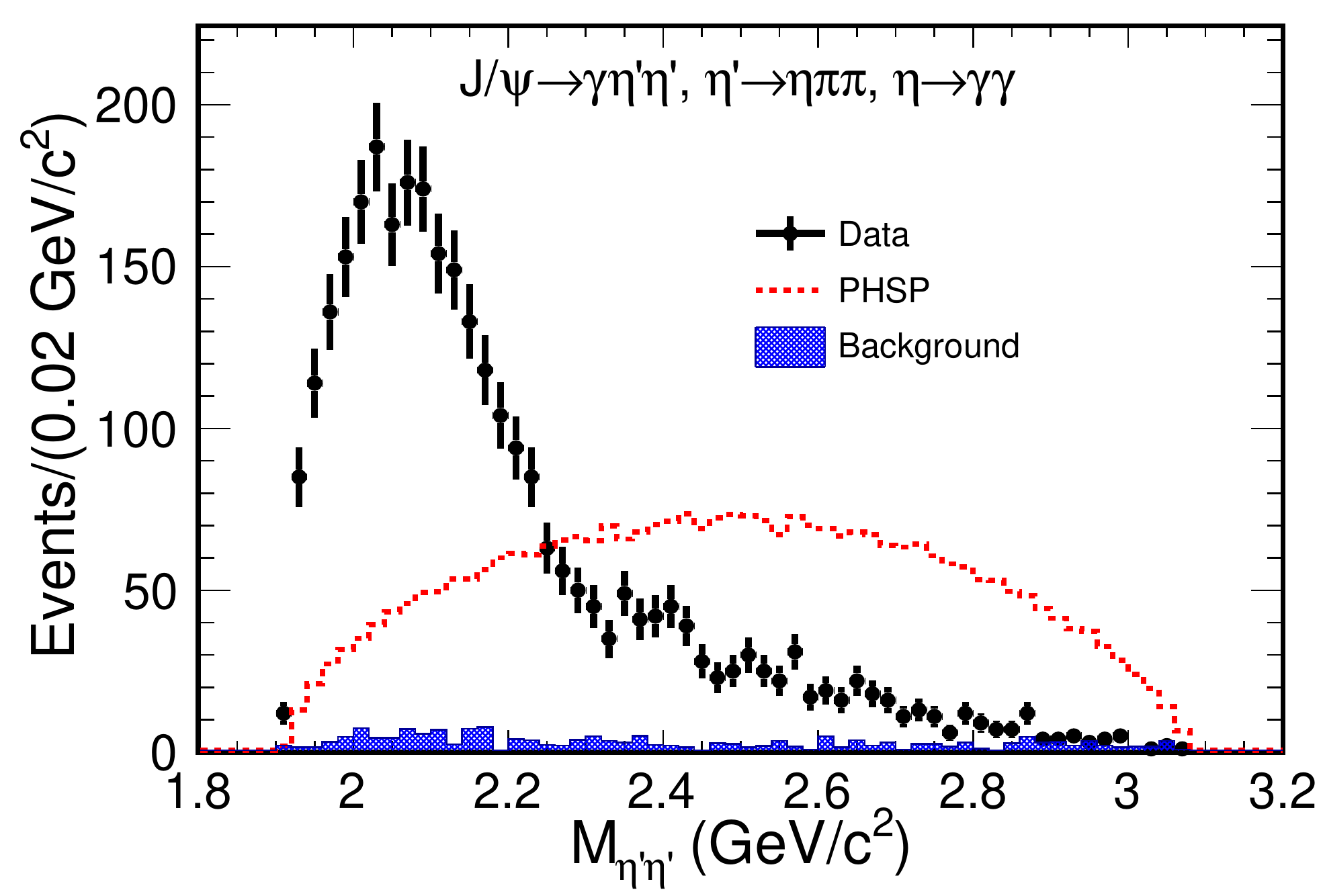} \put(-30,120){(a)} 
	\hspace*{2mm}
 	\includegraphics[width=0.45\textwidth]{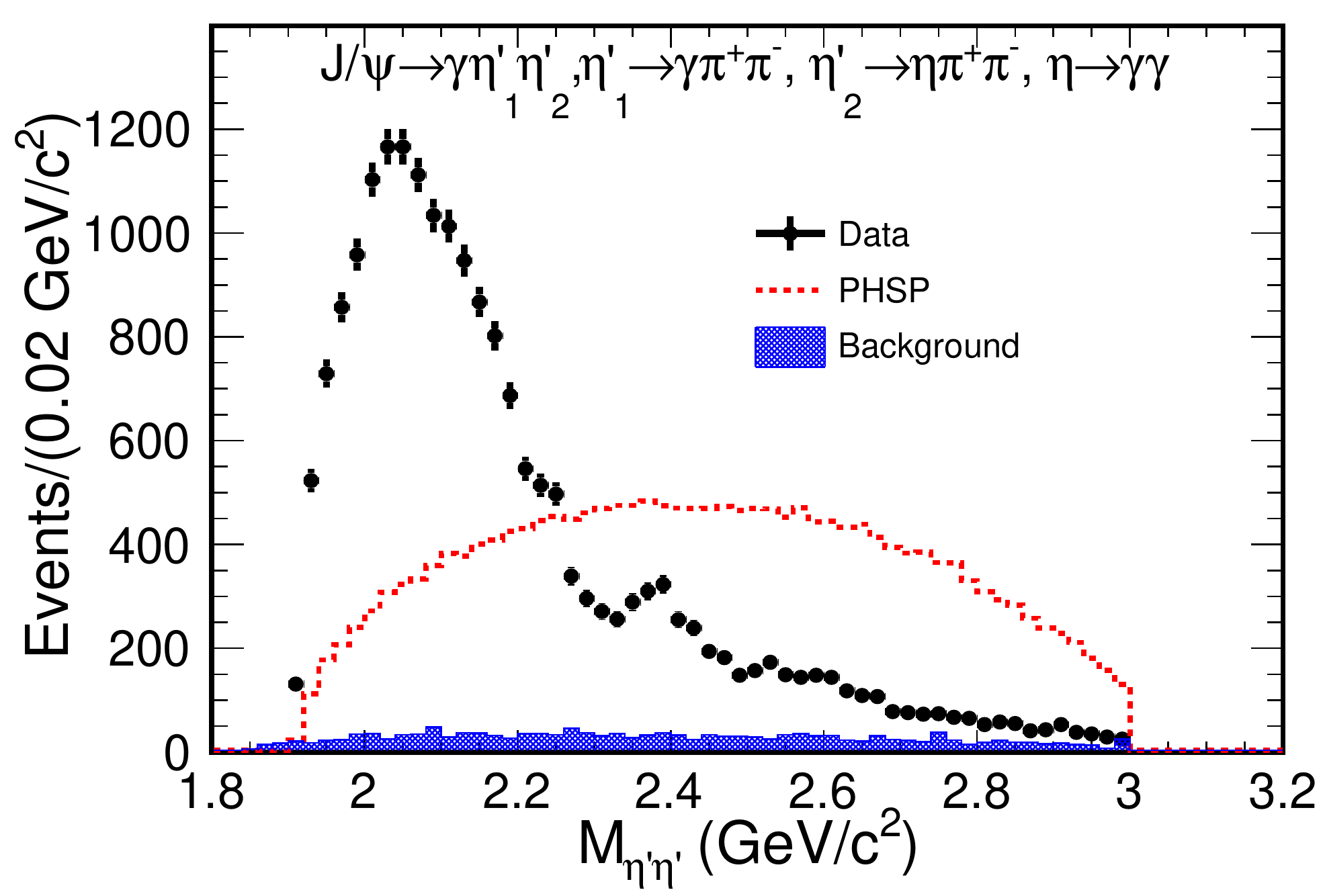} \put(-30,120){(b)}
	\caption{
		\label{fig:mass_bietap}
		Invariant mass distributions of $M_{\etap\etap}$ in (a) mode I and (b) mode II for the selected $\gam\bietap$ candidates.
		The points with error bars and the dashed line show data and the signal PHSP MC samples (normalized by integral), respectively; 
		the shaded histograms show the background contributions	estimated from $\etap\etap$ sidebands.}
\end{figure*}

\begin{figure*}[htbp!]
	\centering
	\includegraphics[width=0.45\textwidth]{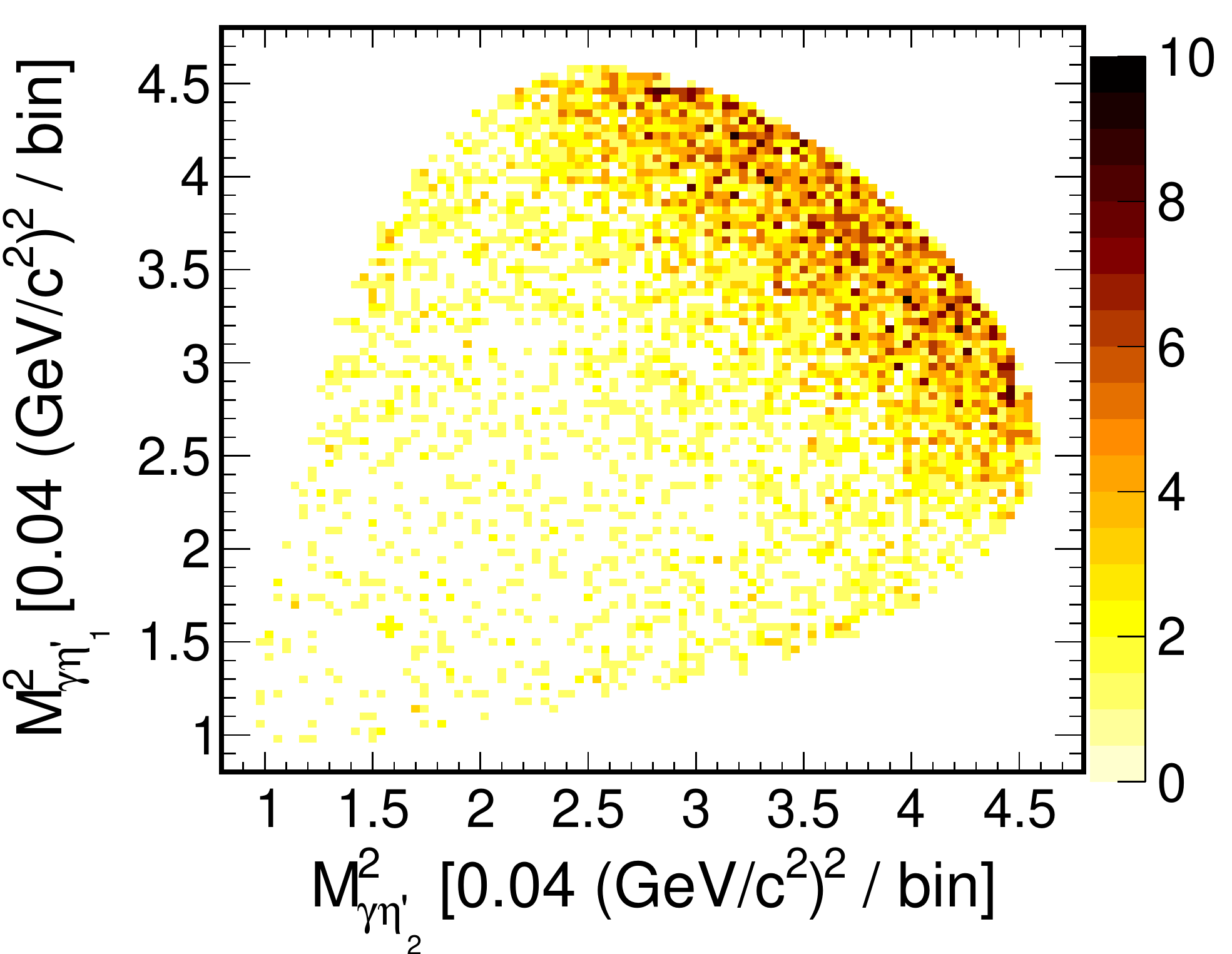} \put(-50,140){(a)}
	\hspace*{3mm}
	\includegraphics[width=0.45\textwidth]{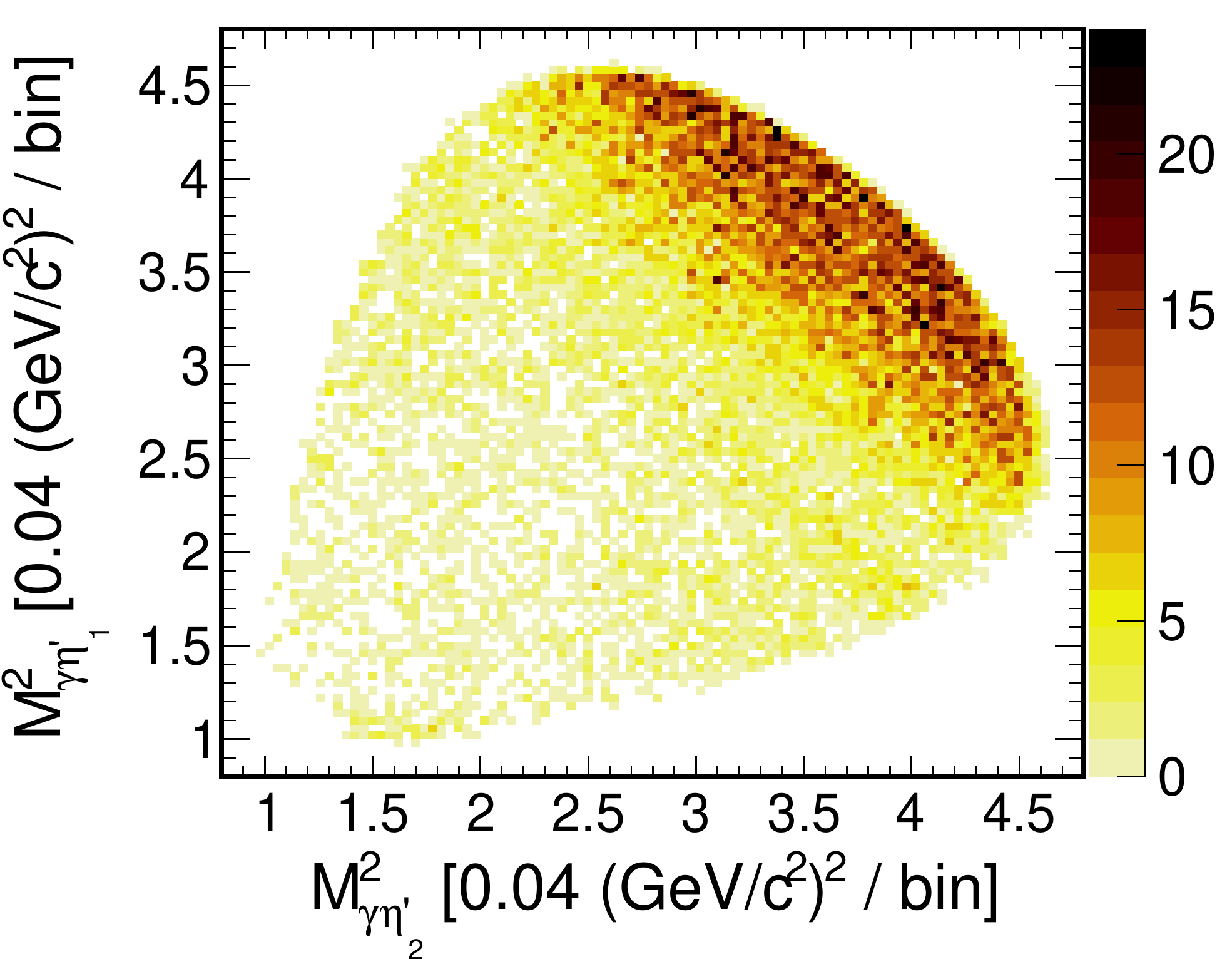} \put(-50,140){(b)}
	\caption{
	\label{fig:dalitz}
	Dalitz plots for the selected candidates in (a) mode I and (b) mode II.}	
\end{figure*}

In mode II, with $\jpsi\to\gamma\etap_1\etap_2,~\etap_1\to\gam\pp,~\etap_2\rightarrow\eta\pp$, and~$\eta\rightarrow\gam\gam$, the final state consists of 4$\gam$ and 2($\pp$).
The subscripts (1, 2) correspond to the different decay modes of $\etap$.
A five-constraint (5C) kinematic fit is performed under the hypothesis of $\jpsi\rightarrow\gam\gam\eta\pp\pp$.
In case of more than one combination, the one with the least $\chi_{\rm 5C}^{2}$ is retained, and the condition $\chi_{\rm 5C}^{2}<$ 75 is required.
To suppress background contributions with one more or one less photon, four-constraint kinematic fits are performed separately under the hypotheses of $\jpsi\rightarrow3\gam\pp\pp,~\jpsi\rightarrow4\gam\pp\pp,~\jpsi\rightarrow5\gam\pp\pp$, whose $\chisq$ values are denoted as $\chi^{2\prime}_{3\gam},~\chi^{2\prime}_{4\gam},~\chi^{2\prime}_{5\gam}$, respectively.
Events with $\chi^{2\prime}_{4\gam}<\chi^{2\prime}_{3\gam}$ and $\chi^{2\prime}_{4\gam}<\chi^{2\prime}_{5\gam}$ are accepted.
To suppress background due to $\pi^{0}\to\gamma\gamma$, $|M_{\gamma\gamma} - m_{\pi^{0}}| >$ 0.02~GeV/$c^{2}$ is required for all possible two-photon combinations excluding the one forming the $\eta$, where $m_{\piz}$ is the nominal mass of the $\piz$.
Invariant mass of the $\pp$ in $\etap\rightarrow\gam\pp$ is required to be within the $\rho$ mass region, 0.4~GeV/$c^{2}< M_{\pi^{+}\pi^{-}} <$ 0.85~GeV/$c^{2}$.
The $\etap$ candidates are formed from the $\gam\pp$ and $\eta\pp$ combination with the least $\frac{\left| M_{\gamma \pp}-m_{\etap} \right|^2}{\sigma^2_{\gamma \pp}} + \frac{\left| M_{\eta\pp}-m_{\etap} \right|^2}{\sigma^2_{\eta \pp}}$, and are then selected with $|M_{\gam\pp} - m_{\etap}| <$ 0.02 GeV/$c^{2},~|M_{\eta\pp} - m_{\etap}| <$ 0.01 GeV/$c^{2}$.
Here, $M_{\gam\pp}$ and $M_{\eta\pp}$ are the invariant masses of $\gam\pp$ and $\eta\pp$, as shown in Figs.~\ref{fig:metap}(b) and~(c); 
$\sigma_{\gamma \pp}$ and $\sigma_{\eta \pp}$ are the invariant mass resolutions of $\gam\pp$ and $\eta\pp$, which are determined to be 6.8~MeV/$c^{2}$ and 3.4~MeV/$c^{2}$ from fitting with a double Gaussian function plus a first order polynomial function to the individual distributions. 
To suppress backgrounds introduced by fake photons, such as $\jpsi\rightarrow\pip\pim\piz\etap$, $M_{\etap\etap}<$ 3.0~GeV/$c^2$ is required.
The distribution of $M_{\gam\pp}$ versus $M_{\eta\pp}$ of the remaining events is shown in Fig.~\ref{fig:sdb_divide}(b), where the red box labeled as ``0'' shows the signal region.

After applying the requirements above, 3081 and 19866 events survived for mode~I,~II, respectively.
The invariant mass distribution of $\etap\etap$ ($M_{\etap\etap}$) and the Dalitz plot for these two modes 
are shown in Figs.~\ref{fig:mass_bietap} and~\ref{fig:dalitz}.

No background events with $\bietap$ in the final state are observed in these two decay modes in a study using a MC sample of $10.01\times10^{9}$ $\jpsi$ inclusive decays with a generic event type analysis tool~\cite{Zhou:2020ksj}.
Non-$\bietap$ backgrounds are estimated using the $\etap\etap$ sideband events from data.
The two-dimensional sidebands for these two modes are illustrated by solid boxes labeled from ``1'' to ``8'' in Fig.~\ref{fig:sdb_divide}, where the sideband regions are defined as 0.02~GeV/$c^{2}$~$\le\left| M_{\eta\pp(1,2)}-m_{\etap} \right|\le$~0.04~GeV/$c^{2}$ for mode I and 0.06~GeV/$c^{2}$~$\le\left| M_{\gam\pp}-m_{\etap} \right|\le$~0.10~GeV/$c^{2}$, 0.03~GeV/$c^{2}$~$\le\left| M_{\eta\pp}-m_{\etap} \right|\le$~0.05~GeV/$c^{2}$ for mode II, respectively.
The shaded histograms in Figs.~\ref{fig:mass_bietap}(a) and~(b) show the background contributions from the normalized sideband events in mode I and mode II, where the numbers of events are 175 and 1576, respectively, corresponding to background fractions of 5.7\% and 7.9\%.

\section{Partial wave analysis}
\subsection{Analysis method}
A PWA is performed to disentangle the structures present in $\jpsi\rightarrow\gam\etap\etap$ decays using the GPUPWA framework~\cite{Berger:2010zza}.
The quasi two-body decay amplitudes in the sequential decay processes $\jpsi\rightarrow\gam X,~X\rightarrow\bietap$ and $\jpsi\rightarrow\etap X,~X\rightarrow\gam\etap$ are constructed using covariant tensor amplitudes described in Ref.~\cite{Zou:2002ar}.
For radiative and hadronic $\jpsi$ decays to mesons, the general forms for covariant tensor amplitudes $A$ are formulated as
\begin{equation}
	A = \psi_{\mu}(m_1)e_{\nu}^{*}(m_2)A^{\mu\nu} = \psi_{\mu}(m_1)e_{\nu}^{*}(m_2)\sum_i\Lambda_{i}U_i^{\mu\nu},
\end{equation}
and 
\begin{equation}
	A = \psi_{\mu}(m)A^{\mu} = \psi_{\mu}(m)\sum_i\Lambda_{i}U_i^{\mu},
\end{equation}
respectively, according to Ref.~\cite{Zou:2002ar}, where $\psi_{\mu}(m_1)$ is the $\jpsi$ polarization four-vector, $e_{\mu}(m_2)$ is the polarization vector of the photon and $U_i^{\mu\nu}$($U_i^{\mu}$) is the partial wave amplitude with a coupling strength determined by the complex parameter $\Lambda_i$.
The partial wave amplitudes $U_i$ for the intermediate states used in the analysis are constructed with the four-momenta of the particles in the final states, and their specific expressions are given in Ref.~\cite{Zou:2002ar}.

Each intermediate resonance $X$ is parametrized by a constant-width, relativistic Breit-Wigner (BW) propagator,
\begin{equation}
	BW(s) = \frac{1}{M^2-s-iM\Gamma},
\end{equation}
where $s$ is the invariant mass squared of $\bietap$ or $\gam\etap$, and $M,~\Gamma$ are the mass and width of the intermediate resonance.

The complex parameters of the amplitudes and resonance parameters are determined by an unbinned maximum likelihood fit.
The probability to observe the $i$-th event characterized by the measurement $\xi$, i.e. the measured four-momenta of the particles in the final states, is
\begin{equation}
	P(\xi_i) = \frac{\omega(\xi_i)\eff(\xi_i)}{\int \text{d}\xi \omega(\xi)\eff(\xi)},
\end{equation}
where $\eff(\xi)$ is the detection efficiency, $\omega(\xi_i)\equiv(\frac{\text{d}\sigma}{\text{d}\Phi})_i$ is the differential cross section, and d$\Phi$ is the standard element of phase space.
The differential cross section is 
\begin{equation}
	\frac{\text{d}\sigma}{\text{d}\Phi} = \left| \sum A(J^{PC}) \right|^2,
\end{equation}
where $A(J^{PC})$ is the amplitude for all possible intermediate resonances with spin-parity $J^{PC}$.
$\int \text{d}\xi \omega(\xi)\eff(\xi)\equiv\sigma^\prime$ is the 
measured yields.

The likelihood for observing the $N$ events in the data sample is 
\begin{equation}
	{\cal L} = \prod^N_{i=1}P(\xi_i)=\prod^N_{i=1}\frac{(\frac{\text{d}\sigma}{\text{d}\Phi})_i\cdot\eff(\xi_i)}{\sigma^\prime}.
\end{equation}

For technical reasons, rather than maximizing ${\cal L}$, $-{\rm ln}~{\cal L}$ is minimized, i.e. 
\begin{equation}
	-{\rm ln}~{\cal L} = -\sum^N_{i=1}{\rm ln}\left(\frac{(\frac{\text{d}\sigma}{\text{d}\Phi})_i}{\sigma^\prime}\right)-\sum^N_{i=1}{\rm ln}~\eff(\xi_i)
\end{equation}
for a given data set. 
The second term is a constant and has no impact on the determination of the parameters or on the related changes of ${\rm -ln}~{\cal L}$, which is then defined as
\begin{equation}
	-{\rm ln}~{\cal L} = -\sum^N_{i=1}{\rm ln}\left(\frac{(\frac{\text{d}\sigma}{\text{d}\Phi})_i}{\sigma^\prime}\right)=-\sum^N_{i=1}{\rm ln}\left(\frac{\text{d}\sigma}{\text{d}\Phi}\right)_i+N~{\rm ln}~\sigma^\prime
\end{equation}
in the fitting.
The free parameters are optimized by using MINUIT~\cite{James:1975dr}.
The measured yields $\sigma^\prime$ is evaluated using MC techniques.
An MC sample of $N_\text{gen}$ is generated with signal events that are uniformly distributed in phase space.
These events are subjected to the event selection criteria and yield a sample of $N_\text{acc}$ accepted events.
The normalization integral is computed as 
\begin{equation}
	\int \text{d}\xi\omega(\xi)\eff(\xi) = \sigma^\prime \rightarrow \frac{1}{N_\text{gen}}\sum_k^{N_\text{acc}}\left(\frac{\text{d}\sigma}{\text{d}\Phi}\right)_k.
\end{equation}
To take into account the non-$\bietap$ background contribution in data, the negative log-likelihood (NLL) value obtained from events in the $\bietap$ signal region is subtracted by the NLL value obtained from events in the $\bietap$ sideband regions, i.e., 
\begin{equation}
	-{\rm ln}~{\cal L}_\text{sig} = -\left({\rm ln}~{\cal L}_\text{data} - \sum_i\omega_i\cdot{\rm ln}~{\cal L}_{\text{bkg}_i} \right),
\end{equation}
where $\omega_i$ represents the scaling factor of background events in different $\bietap$ sideband region, and ${\cal L}_{\text{bkg}_i}$ means the likelihood from corresponding background events.

The number of fitted events $N_X$ for an intermediate resonance $X$, which has $N_{W_{X}}$ independent partial wave amplitudes $A_j$, is defined as 
\begin{equation}
	N_X= \frac{\sigma_X}{\sigma^\prime}\cdot N^\prime,
\end{equation}
where $N^\prime$ is the number of selected events after background subtraction and 
\begin{equation}
	\sigma_X = \frac{1}{N_\text{gen}}\sum_k^{N_\text{acc}}\left| \sum_j^{N_{W_X}}(A_j)_k \right|^2
\end{equation}
is the measured cross section of the resonance $X$ and is calculated with the same MC sample as the measured total cross section $\sigma^\prime$.
The detection efficiency $\eff_X$ is obtained by the partial wave amplitude weighted MC sample,
\begin{equation}
	\eff_X = \frac{\sigma_X}{\sigma_X^\text{gen}}=\frac{\sum_k^{N_\text{acc}}\left| \sum_j^{N_{W_X}}(A_j)_k\right|^2}{\sum_i^{N_\text{gen}}\left| \sum_j^{N_{W_X}}(A_j)_i\right|^2}.
\end{equation}

Based on GPUPWA, a combined fit for mode I and II is performed.
The combined branching fraction of $\jpsi\rightarrow\gam X,~X\rightarrow\bietap$ or $\jpsi\rightarrow\etap X,~X\rightarrow\gam\etap$ is calculated by 
\begin{widetext}
\begin{equation}
\label{eq:comb_br}
	{\cal B}(\jpsi\rightarrow\gam X~{\rm or}~\etap X \rightarrow\gam\bietap)=\frac{N_{X_{\text{I}}}+N_{X_{\text{II}}}}{N_{\jpsi}\cdot {\cal B}_{\etap\rightarrow\eta\pp}\cdot {\cal B}_{\eta\rightarrow\gam\gam}\cdot ({\cal B}_{\etap\rightarrow\eta\pp}\cdot {\cal B}_{\eta\rightarrow\gam\gam}\cdot \eff_{X_{\text{I}}}+{\cal B}_{\etap\rightarrow\gam\pp}\cdot\eff_{X_{\text{II}}}\cdot 2)},
\end{equation}
\end{widetext}
where the subscripts (I, II) correspond to mode I and mode II, respectively,
$N_{\jpsi}$ is the total number of $\jpsi$ events, and ${\cal B}_{\etap\rightarrow\eta\pp},~{\cal B}_{\etap\rightarrow\gam\pp},~{\cal B}_{\eta\rightarrow\gam\gam}$ are the branching fractions of $\etap\rightarrow\eta\pp,~\etap\rightarrow\gam\pp,~\eta\rightarrow\gam\gam$ from Ref.~\cite{Zyla:2020zbs}, respectively.
The factor 2 in the second term of the denominator in Eq.~\ref{eq:comb_br} represents the combination of the two decay modes of $\etap$.

\subsection{PWA results} \label{sub:pwa_result}
\begin{figure*}[htbp]
	\centering
	\includegraphics[width=0.45\textwidth]{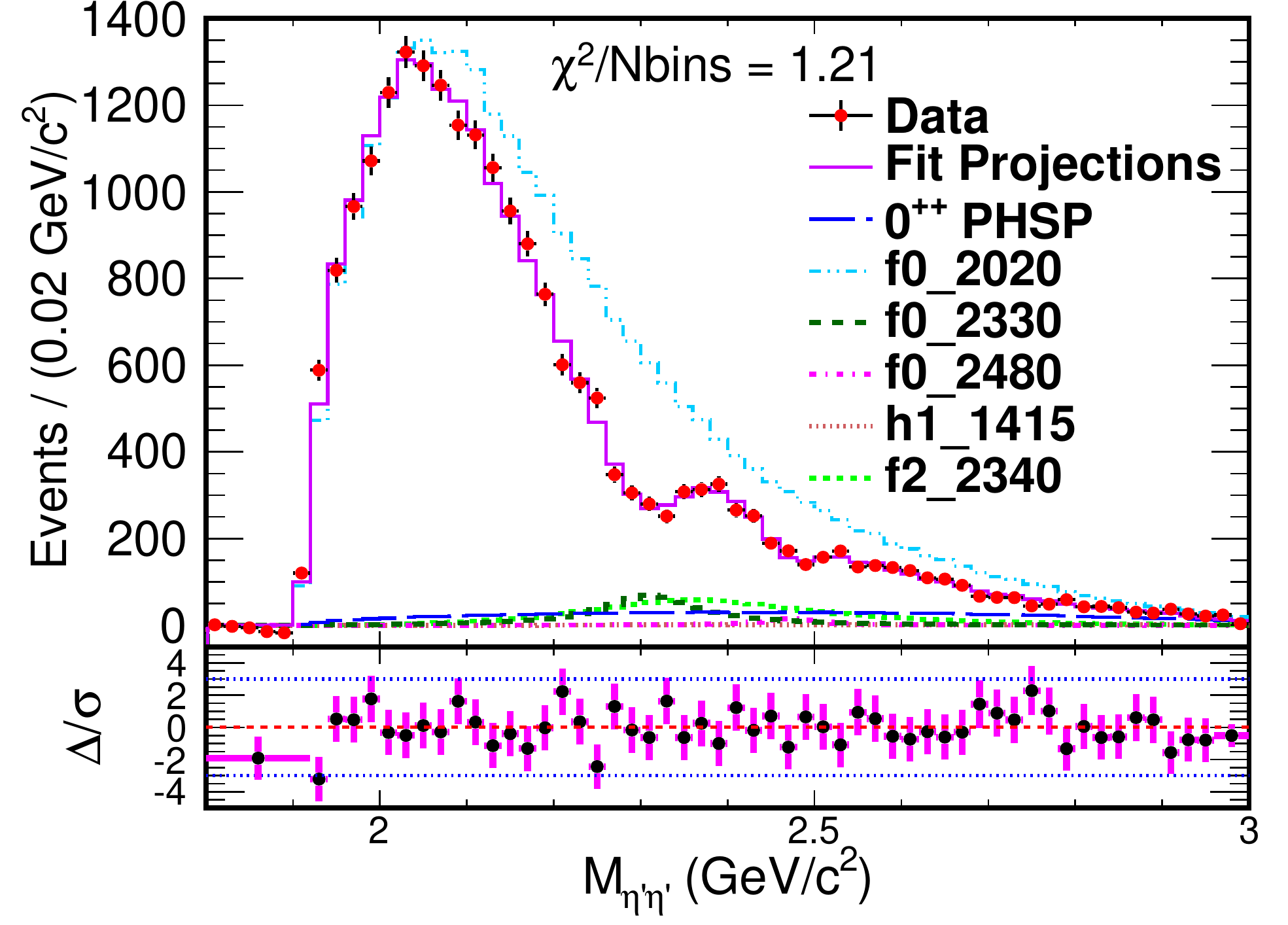} 
	\put(-28,152){(a)}
	\hspace*{2mm}
	\includegraphics[width=0.45\textwidth]{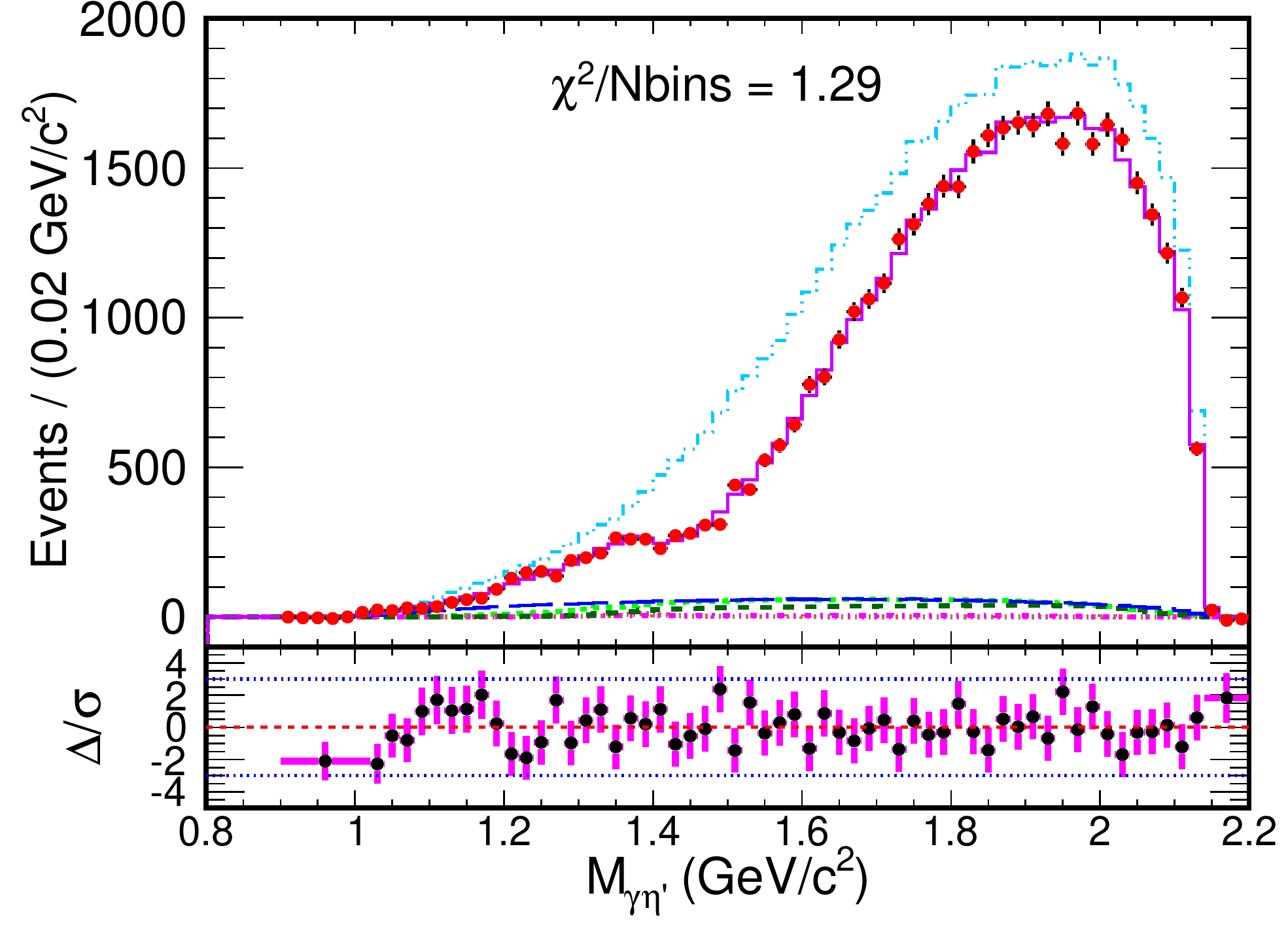}
	\put(-28,152){(b)} 
	\\
	\vspace*{2mm}
	\includegraphics[width=0.45\textwidth]{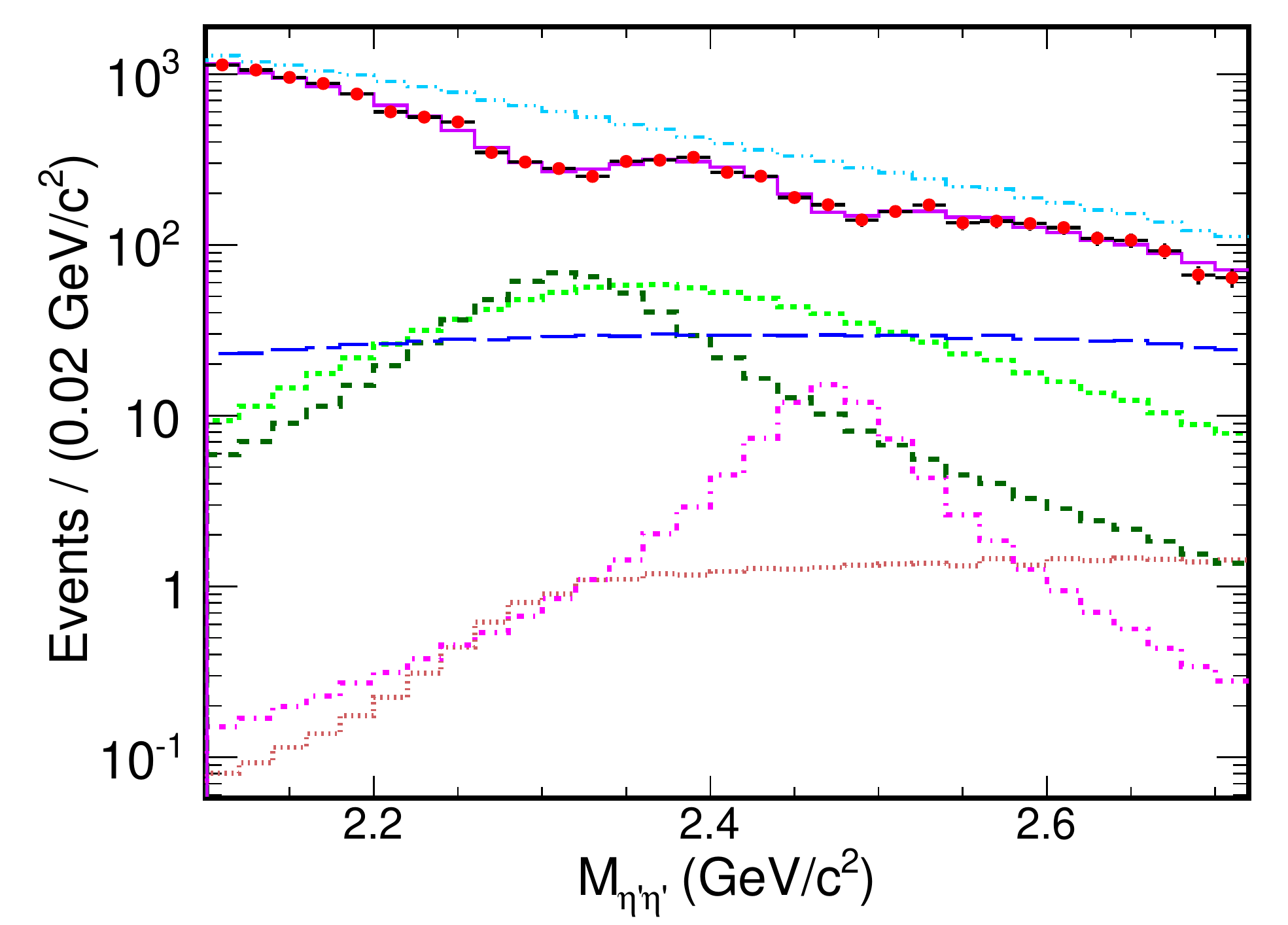}
	\put(-28,135){(c)} 
	\hspace*{2mm}	
	\includegraphics[width=0.45\textwidth]{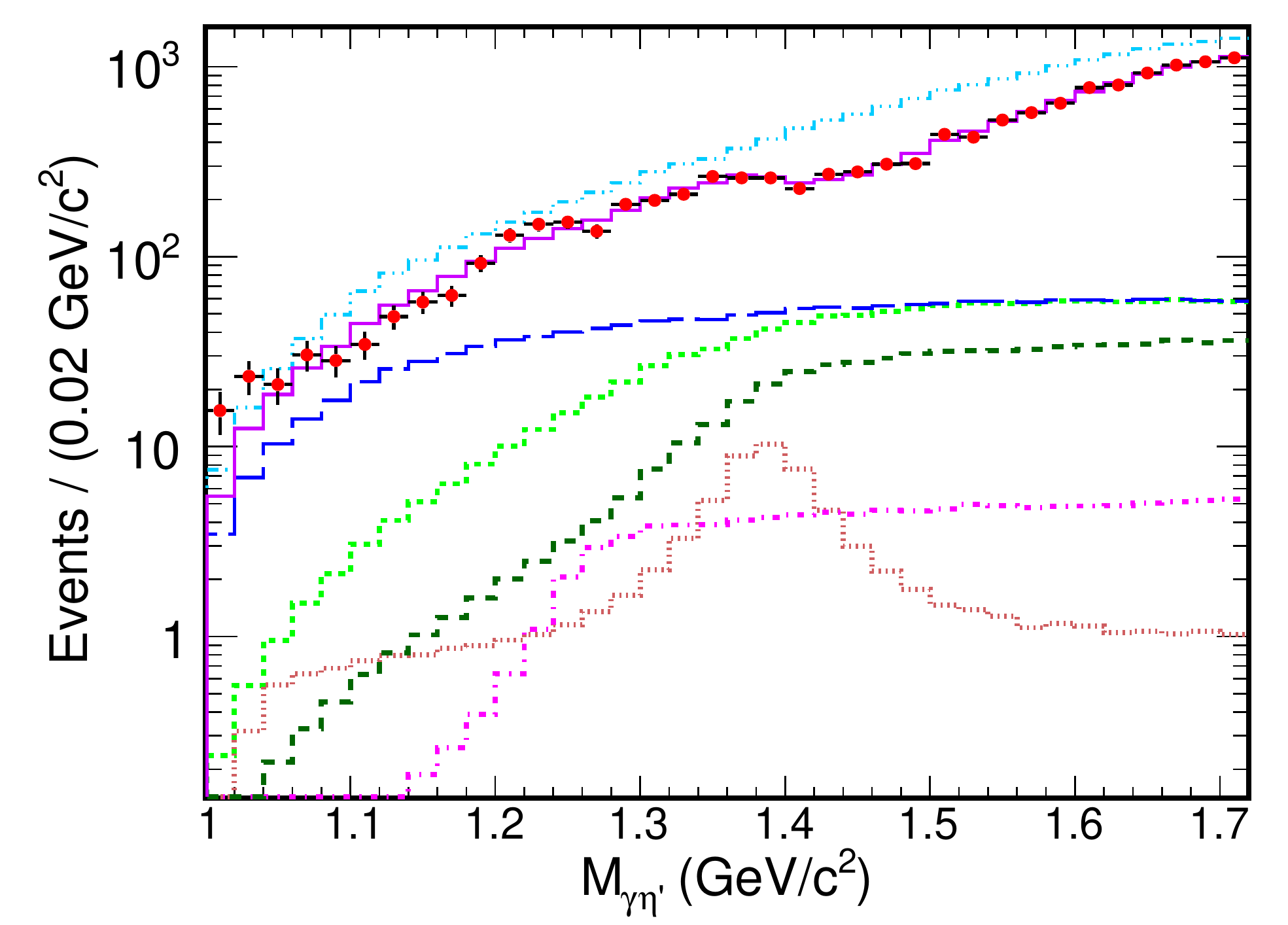}
	\put(-28,135){(d)} 
	\\
	\vspace*{2mm}
	\includegraphics[width=0.45\textwidth]{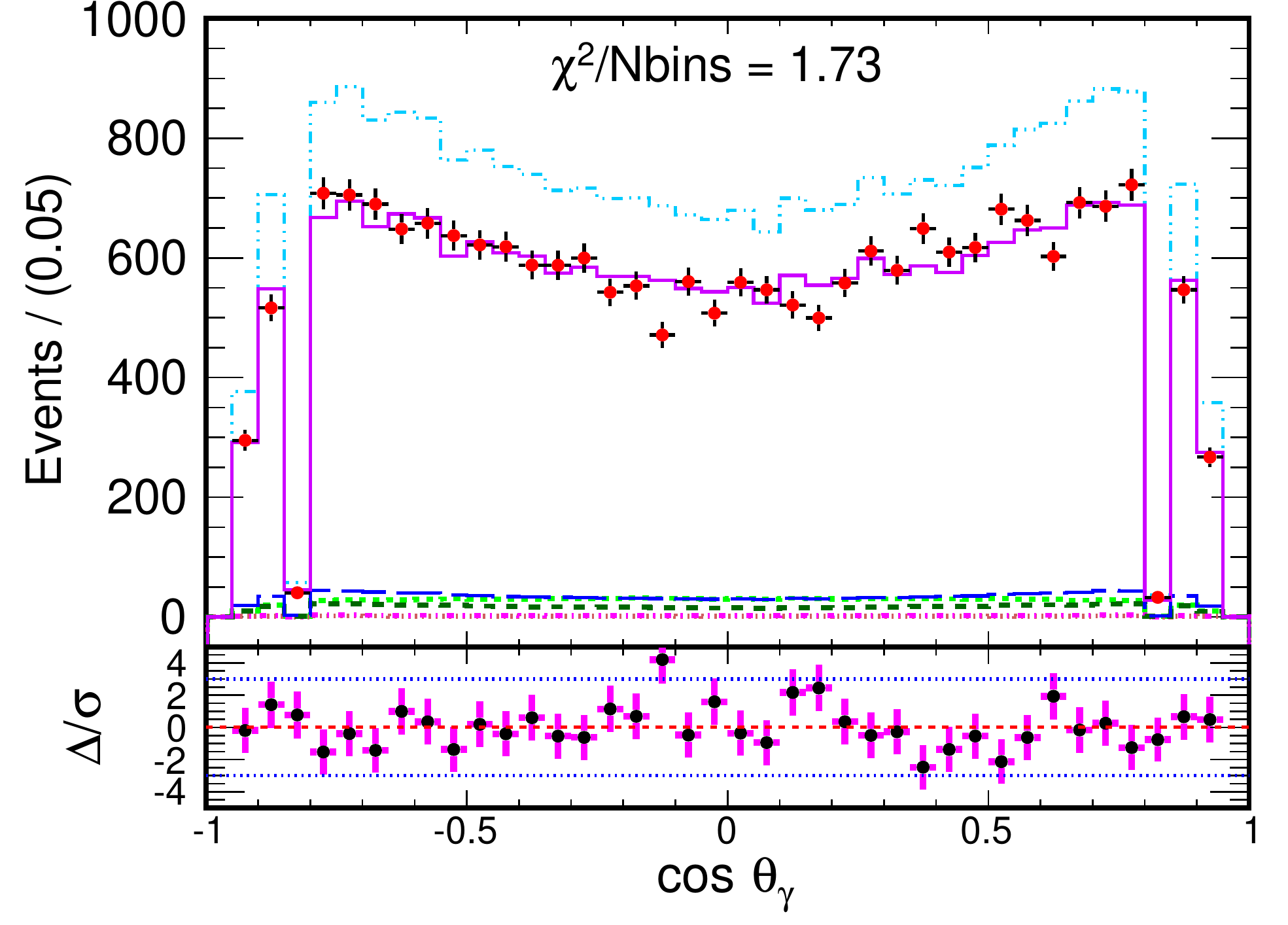}
	\put(-28,152){(e)} 
	\hspace*{2mm}	
	\includegraphics[width=0.45\textwidth]{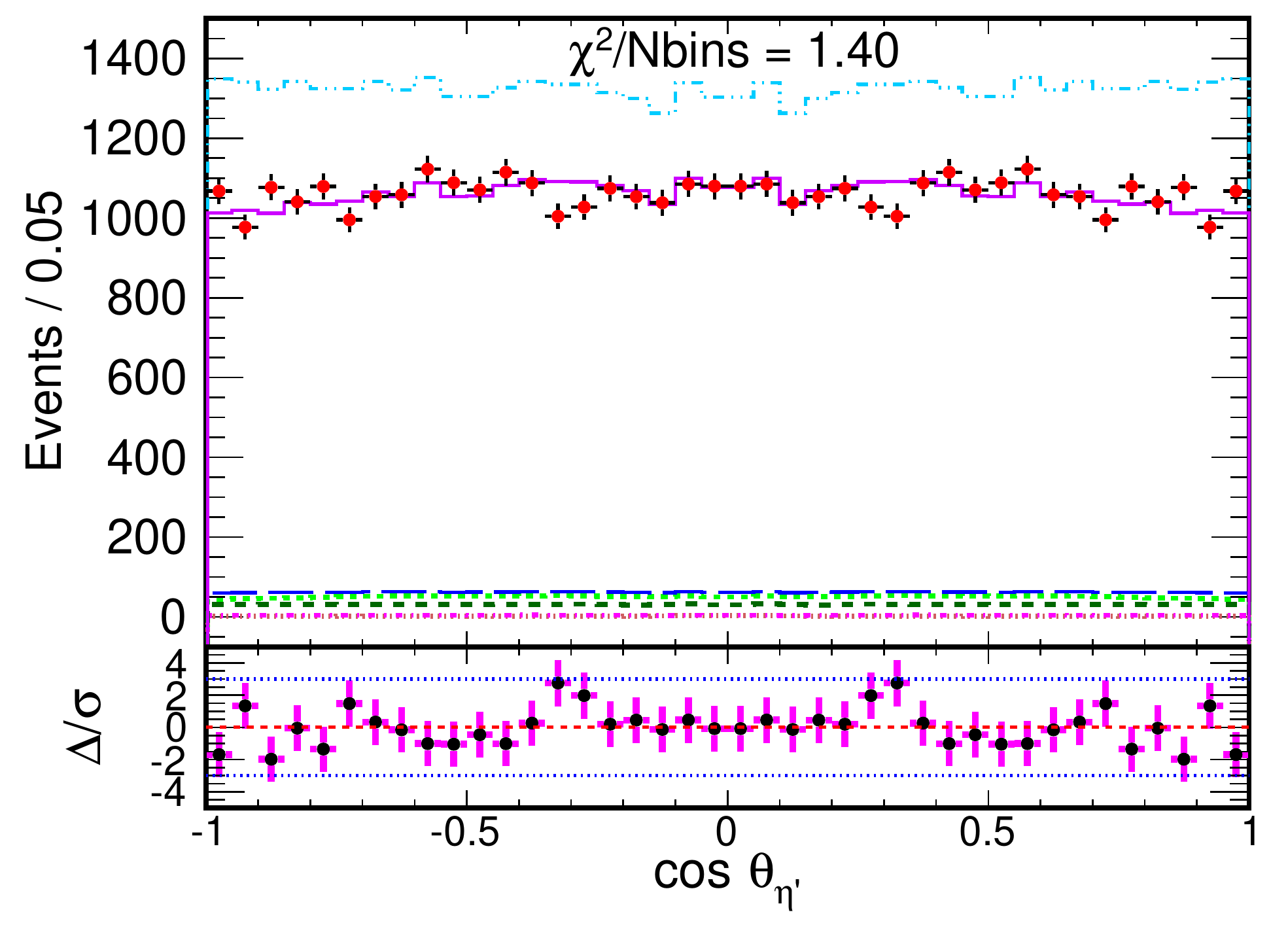}
	\put(-28,152){(f)} 
	\\
	\vspace*{2mm}
	\includegraphics[width=0.45\textwidth]{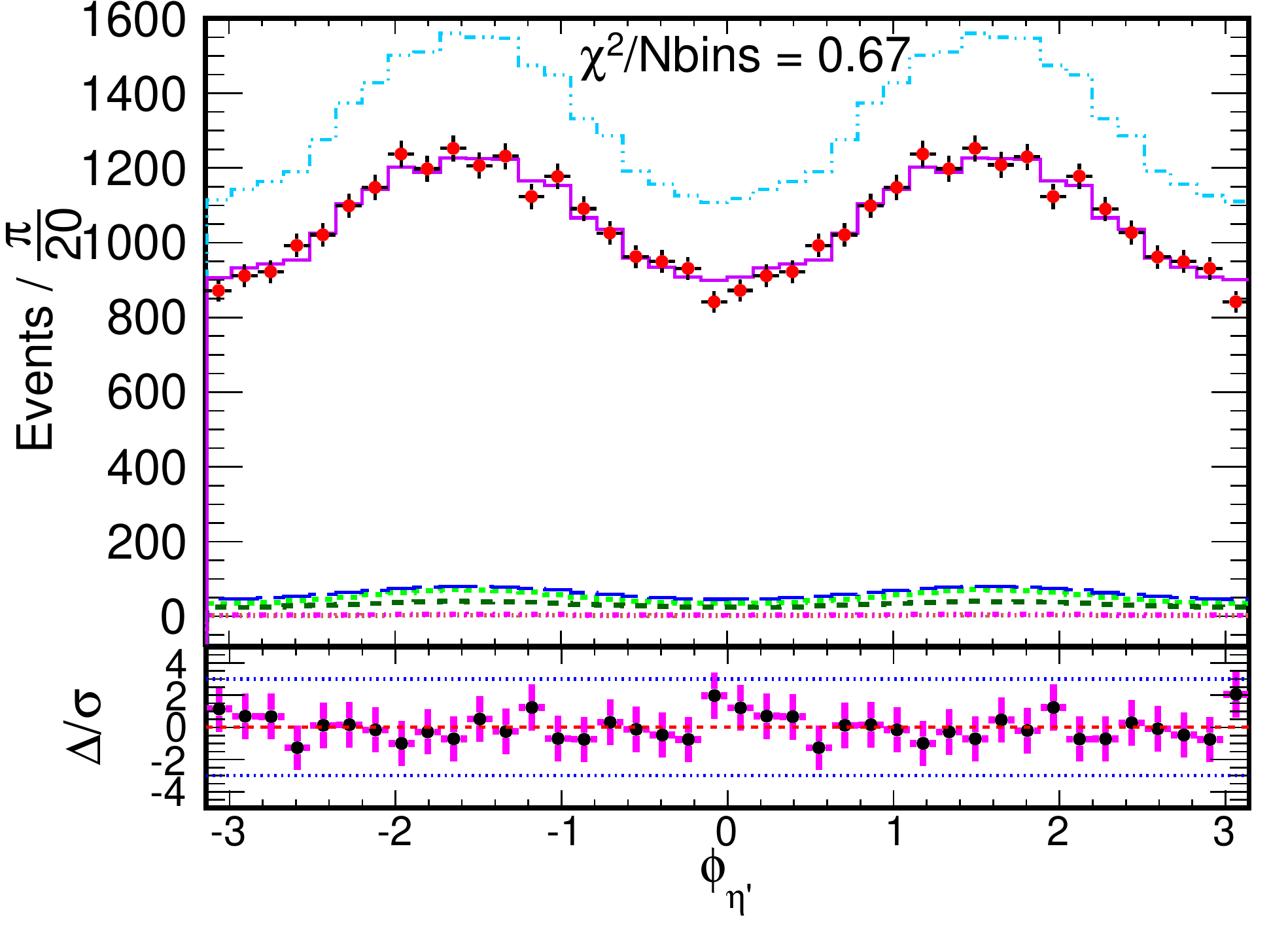}
	\put(-28,152){(g)}
	\caption{\label{fig:comb_fit_result}
	Superposition of data and the combined PWA fit projections for: Invariant mass distributions of (a) $M_{\bietap}$, (b) $M_{\gam\etap}$ (the bins at the left and right edges in the pull distributions are combined due to low statistics); zoomed view of invariant mass distributions of (c) $M_{\bietap}$, (d) $M_{\gam\etap}$ in logarithm scale; cos$~\theta$ of (e) $\gamma$ in the $\jpsi$ rest frame, (f) $\etap$ in the $\bietap$ rest frame; (g) azimuthal angle between the normals to the two decay planes of $\etap$ in the $\bietap$ rest frame.
	}
\end{figure*}

In this analysis, all combinations of possible
resonances in the PDG~\cite{Zyla:2020zbs} and Ref.~\cite{Bugg:2004xu} with $J^{PC}=0^{++},~2^{++},~4^{++}$ in $\bietap$ and $J^{PC}=1^{+-},~1^{--}$
in $\gam\etap$ are considered, as listed in Tab.~\ref{tab:resonances}.
Because of centrifugal barriers, production of $\gam\etap$ intermediate states with $J\ge 3$ and of $\bietap$ intermediate states with $J\ge 6$ in $\jpsi\rightarrow\gam\bietap$ is not considered.
Changes in the NLL value and the number of free parameters in the fit with and without a resonance are used to evaluate its statistical significance.
All components with significance greater than 5$\sigma$ are retained in the baseline model.
The baseline model contains two $0^{++}$ resonances ($f_0(2020),~f_0(2330)$), one $2^{++}$ resonance ($f_2(2340)$), one $1^{+-}$ resonance ($h_1(1415)$), and the non-resonant decay of $\jpsi\rightarrow\gam\bietap$, which is modeled by a $0^{++}$ phase space distribution ($0^{++}$ PHSP) of the $\bietap$ system.
In the high energy region of the $\bietap$ spectrum, a new scalar, $f_0(2480)$, is needed to describe 
data with a significance of 6.0$\sigma$.
By taking into account the look-elsewhere effect~\cite{Gross:2010qma}, the significance is 5.2$\sigma$.
Additional resonances listed in Tab.~\ref{tab:resonances} are tested.
None of them has a statistical significance larger than 5$\sigma$.
The existence of possible additional resonances is further studied by performing scans for extra resonances ($J^{PC}=0^{++},~2^{++},~4^{++},~1^{+-},~1^{--}$) with different masses and widths.
The scan results yield no evidence for extra intermediate states.
The masses and widths of all resonances in the baseline model, product branching fractions of $\jpsi\rightarrow\gam X,~X\rightarrow\bietap$ or $\jpsi\rightarrow\etap X,~X\rightarrow\gam\etap$, and the statistical significances are summarized in Tab.~\ref{tab:base_solu}, where the first uncertainties are statistical and the second are systematic.
The fit fraction of each component and their interference fractions are listed in Tab.~\ref{tab:frac_fit}.
The comparisons of data and the PWA fit projection (weighted by MC efficiencies) of the invariant mass distributions of $\bietap$ and $\gam\etap$ for the fitted parameters are shown in Figs.~\ref{fig:comb_fit_result}(a) and (b).
To make the components with small fit fractions (the $f_0(2480)$ and the $h_1(1415)$) visible, the zoomed view of these two distributions in logarithm scale are shown in Figs.~\ref{fig:comb_fit_result}(c) and (d).
The comparisons of the projected data and MC angular distributions are shown in Figs.~\ref{fig:comb_fit_result}(e),~(f) and~(g).
The $\chisq/N_{\rm bins}$ value is displayed on each figure to demonstrate the goodness of fit, where $N_{\rm bins}$ is the number of bins of each figure and the $\chisq$ is defined as:
\begin{equation}
	\chisq=\sum^{N_{\rm bins}}_{i=1}\frac{\left(n_i-\nu_i\right)^2}{\nu_i},
\end{equation}
where $n_i$ and $\nu_i$ are the number of events for the data and the fit projections with the baseline model in the $i$-th bin of each figure, respectively.

\begin{table}[htbp!]
\centering
\caption{States considered in this analysis with $J^{PC}=0^{++},~2^{++},~4^{++}$ in $\bietap$ and $J^{PC}=1^{+-},~1^{--}$ in $\gam\etap$ and the significances of additional resonances. ``*'' means the state is from Ref.~\cite{Bugg:2004xu}. PHSP means the non-resonant contribution.}
\label{tab:resonances}
   \renewcommand\arraystretch{1.2}
    \begin{tabular}{c |c c c }
    \hline \hline
     & State & $J^{PC}$ & Sig.($\sigma$) \\ \hline    
    \multirow{5}{*}{Baseline model} & $0^{++}$ PHSP & $0^{++}$ & ...	  \\ \cline{2-4}
    ~ & $f_{0}(2020)$ & $0^{++}$ & ...	\\ \cline{2-4}
    ~ & $f_{0}(2330)$  & $0^{++}$ & ...	\\ \cline{2-4}
    ~ & $f_{2}(2340)$ &  $2^{++}$ &  ...	    \\ \cline{2-4}
    ~ & $h_{1}(1415)$	&	$1^{+-}$	&	...	\\	\hline
    \multirow{38}{*}{Additional states} & 
    	$f_{0}(2020)^*$ & $0^{++}$ & 0.0 	\\ \cline{2-4}
    ~ & $f_{0}(2060)$ & $0^{++}$ & 	0.0		\\ \cline{2-4}
    ~ & $f_{0}(2100)$ & $0^{++}$ & 	0.0	\\ \cline{2-4}
    ~ & $f_{0}(2102)^*$ & $0^{++}$ & 	0.0	\\ \cline{2-4}
    ~ & $f_{0}(2200)$      & $0^{++}$ & 0.1	\\ \cline{2-4}
    ~ & $2^{++}$ PHSP & $2^{++}$ & 	1.1	\\ \cline{2-4}
    ~ & $f_{2}(1910)$ & $2^{++}$ & 	0.9		\\ \cline{2-4}
	~ & $f_{2}(1934)^*$ & $2^{++}$ & 0.4	\\ \cline{2-4}
	~ & $f_{2}(1950)$ & $2^{++}$ & 	0.2	\\ \cline{2-4}
	~ & $f_{2}(2000)$ & $2^{++}$ & 	0.4	\\ \cline{2-4}
	~ & $f_{2}(2010)$ & $2^{++}$ & 	0.9	\\ \cline{2-4}
	~ & $f_{2}(2010)^*$ & $2^{++}$ & 0.0	\\ \cline{2-4}
	~ & $f_{2}(2140)$ & $2^{++}$ & 	0.7	\\ \cline{2-4}
    ~ & $f_{2}(2150)$ &  $2^{++}$  &  0.0	\\ \cline{2-4}
    ~ & $f_{2}(2220)$ &  $2^{++}$  &  1.4	\\ \cline{2-4}
    ~ & $f_{2}(2240)$ &  $2^{++}$  &  0.2 \\ \cline{2-4}
    ~ & $f_{2}(2295)$ &  $2^{++}$  &  0.3 \\ \cline{2-4}
    ~ & $f_{2}(2300)$ &  $2^{++}$  &  0.9 \\ \cline{2-4}
    ~ & $4^{++}$ PHSP 		&	$4^{++}$ & 0.7 \\ \cline{2-4}
	~ & $f_{4}(2018)^*$		&	$4^{++}$ & 0.2 \\ \cline{2-4}
	~ & $f_{4}(2050)$		&	$4^{++}$ & 0.2 \\ \cline{2-4}
	~ & $f_{4}(2220)$		&	$4^{++}$ & 0.0 \\ \cline{2-4}
	~ & $X(2260)$       	&	$4^{++}$ & 0.5 \\ \cline{2-4}
	~ & $f_{4}(2283)^*$ 	&	$4^{++}$ & 0.6 \\ \cline{2-4}
	~ & $f_{4}(2300)$ 		&	$4^{++}$ & 0.5 \\ \cline{2-4}
	~ & $1^{+-}$ PHSP	&	$1^{+-}$	&  1.8 \\	\cline{2-4}
	~ & $h_{1}(1170)$	&	$1^{+-}$	&  2.3	\\	\cline{2-4}	
	~ & $h_{1}(1595)$	&	$1^{+-}$	&  0.2	\\	\cline{2-4}
	~ & $1^{--}$ PHSP 	&	$1^{--}$	&  0.1	\\ \cline{2-4}
	~ & $\omega(1420)$ 	&	$1^{--}$	&  0.5	 \\ \cline{2-4}
	~ & $\omega(1650)$ 	&	$1^{--}$	& 0.1	\\ \cline{2-4}
	~ & $\phi(1680)$ 	&	$1^{--}$	& 0.4	\\ \cline{2-4}
	~ & $\phi(2170)$ 	&	$1^{--}$	& 2.0	\\ \cline{2-4}
	~ & $\rho(1450)$ 	&	$1^{--}$	& 0.0	\\ \cline{2-4}
	~ & $\rho(1570)$	&	$1^{--}$	& 0.6	\\ \cline{2-4}
	~ & $\rho(1700)$  	&	$1^{--}$	& 0.2 \\ \cline{2-4}
	~ & $\rho(1900)$	&	$1^{--}$	& 2.4 \\ \cline{2-4}
	~ & $\rho(2150)$	&	$1^{--}$	& 0.9 \\ \cline{2-4}
    \hline \hline
    \end{tabular}
\end{table}

\begin{table*}[htbp!] 
\renewcommand\arraystretch{1.5}
	\begin{center}
		{
		\caption{Mass, width, {$\cal B$}($\jpsi\rightarrow\gam X\rightarrow\gam\bietap$) or {$\cal B$}($\jpsi\rightarrow\etap X\rightarrow\gam\bietap$) (B.F.) and significance of each component in the baseline model. The first uncertainties are statistical and the second are systematic.}
		\label{tab:base_solu}
		}
		\begin{tabular}{lcccc}
		\hline \hline		
		Resonance & M(MeV/$c^2$) & $\Gamma$(MeV) & B.F. & Sig.($\sigma$) \\ \hline	
		$f_0(2020)$ & $1982\pm3^{+54}_{-0}$ & $436\pm4^{+46}_{-49}$ & $(2.63\pm0.06^{+0.31}_{-0.46})\times10^{-4}$ & $\gg25$ \\
		$f_0(2330)$ & $2312\pm2^{+10}_{-0}$ & $134\pm5^{+30}_{-9}$ & $(6.09\pm0.64^{+4.00}_{-1.68})\times10^{-6}$ & 16.3 \\
		$f_0(2480)$ & $2470\pm4^{+4}_{-6}$ & $75\pm9^{+11}_{-8}$ & $(8.18\pm1.77^{+3.73}_{-2.23})\times10^{-7}$ & 5.2
		\\
		$h_1(1415)$ & $1384\pm6 ^{+9}_{-0}$ & $66\pm10^{+12}_{-10}$ & $(4.69\pm0.80^{+0.74}_{-1.82})\times10^{-7}$ & 5.3 \\
		$f_2(2340)$ & $2346\pm8^{+22}_{-6}$ & $332\pm14^{+26}_{-12}$ & $(8.67\pm0.70^{+0.61}_{-1.67})\times10^{-6}$ & 16.1 \\
		$0^{++}$ PHSP & ... & ... & $(1.17\pm0.23^{+4.09}_{-0.70})\times10^{-5}$ & 15.7 \\		
		\hline \hline		
		\end{tabular}
	\end{center}
\end{table*}

\begin{table*}[htbp!]
\renewcommand\arraystretch{1.5}

	\begin{center}
		{
		\caption{Fraction of each component or interference fractions between two components (\%) in the baseline model. The uncertainties are statistical only.}
		\label{tab:frac_fit}
		}
		\begin{tabular}{lcccccc}
		\hline \hline
		Resonance & $f_0(2020)$ & $f_0(2330)$ & $f_0(2480)$ & $h_1(1415)$ & $f_2(2340)$ & $0^{++}$ PHSP \\ \hline
		$f_0(2020)$ & $125.0\pm2.7$ & $-23.7\pm1.4$ & $-3.3\pm0.6$ & $0.8\pm0.3$ & $0.5\pm0.2$ & $-14.3\pm1.6$ \\
		$f_0(2330)$ &  & $3.0\pm0.3$ & $-0.1\pm0.1$ & $-0.2\pm0.0$ & $-0.4\pm0.0$ & $2.9\pm0.2$  \\
		$f_0(2480)$ &  &  & $0.4\pm0.1$ & $0.0\pm0.0$ & $0.1\pm0.0$ & $0.0\pm0.1$  \\
		$h_1(1415)$ & & & & $0.2\pm0.0$ & $0.0\pm0.1$ & $-1.0\pm0.1$ \\
		$f_2(2340)$ & & & & & $5.0\pm0.4$ & $-0.6\pm0.1$ \\
		$0^{++}$ PHSP & & & & & & $5.8\pm1.1$ \\
		\hline \hline		
		\end{tabular}
	\end{center}
\end{table*}

Various checks are performed to verify the reliability of the PWA baseline model.
The statistical significance of additional non-resonant contributions with $J^{PC}=2^{++}$, $4^{++}$, $1^{+-}$ or $1^{--}$ is less than 5$\sigma$.
If $0^{++}$ PHSP is replaced by $2^{++}$ or $4^{++}$ PHSP, the NLL value will be worsened by 116.8 or 123.4. 
For the scalar around 2.1~GeV/$c^{2}$, the fitted mass and width are close to the resonance parameters of the $f_0(2020)$ from the PDG~\cite{Zyla:2020zbs}.
If its mass and width are fixed to the resonance parameters of $f_0(2100)$ or $f_0(2200)$ in the PDG, the NLL value will be worsened by 324.1 or 1437.2, respectively. 
Since the $f_0(2020)$ is near the $\bietap{}$ threshold and $\bietap$ is not its dominant decay channel, a mass-dependent-width BW form is used to parametrize the resonance as the following
\begin{gather}
	\label{eq:bw}
	BW(s) = \frac{1}{M^2-s-i\sqrt{s}\cdot\Gamma(s)},
\end{gather}
with
\begin{gather}
	\label{eq:gamma}
	\Gamma(s)=R\cdot\Gamma\left(M^2\right)\left(\frac{M^2}{s}\right)\left(\frac{\rho(s)}{\rho(M^2)}\right)^{2l+1}+(1-R)\cdot\Gamma_{0},
\end{gather}
where the first term of $\Gamma(s)$ corresponds to the decay of $f_0(2020)\rightarrow\bietap{}$, $R$ is the ${\cal B}\left(f_0(2020)\rightarrow\bietap\right)$ which is estimated to be 0.1 from the PDG~\cite{Zyla:2020zbs}, $\rho(s)$ is the momentum of the $\etap$ in the resonance rest frame, $\Gamma_0$ is the constant width of the $f_0(2020)$ from the PDG~\cite{Zyla:2020zbs}.
The difference between the result in the parameterization using a constant-width BW and that of the mass-dependent-width BW is assigned as part of the systematic uncertainties shown in Sec.~\ref{sec:sys}.
If the $J^{PC}$ of the $f_0(2480)$ is changed from $0^{++}$ to $2^{++}$ or $4^{++}$, the NLL value will be worsened by 20.3 or 30.2, respectively.
The significance of the $f_0(2480)(0^{++})$ hypothesis is further examined using a hypothesis test, in which the alternative hypothesis is our baseline model with an additional $f_J(2480)(J^{PC}= 2^{++}~\text{or}~4^{++})$ state.
The changes of NLL when the $f_0(2480)$ is removed from the alternative hypothesis are 25.4 and 30.0, respectively.
The significances of the $0^{++}$ hypothesis over the alternative $J^{PC}=2^{++},~4^{++}$ possibilities are then determined to be 6.3$\sigma$ and 7.0$\sigma$.

\section{Systematic uncertainties} \label{sec:sys}
The sources of systematic uncertainty are divided into two categories.
The first includes the systematic uncertainties from the number of $\jpsi$ events (0.4\%~\cite{Ablikim:2016fal}), MDC tracking (1.0\% each for four charged tracks~\cite{Ablikim:2011es}), pion PID (1.0\% each for four pions~\cite{Ablikim:2011kv}), photon detection efficiency (1.0\% each~\cite{Ablikim:2010zn} for five, four photons in mode I and mode II), kinematic fit~\cite{Ablikim:2012pg} (1.9\%, 0.8\% for mode I and mode II), $\etap$ mass resolution (0.1\%, 0.4\% for mode I and mode II), and ${\cal B}_{\etap\rightarrow\eta\pp}$ (0.5\% for each $\etap(\eta\pp)$), ${\cal B}_{\etap\rightarrow\gam\pp}$ (0.4\% for each $\etap(\gam\pp)$), ${\cal B}_{\eta\rightarrow\gam\gam}$ (0.2\% for each $\eta$)~\cite{Zyla:2020zbs}.
These systematic uncertainties are applied separately to the branching fractions for mode I and mode II, and summarized in Tab.~\ref{tab:1st_sys_err}.
The measurements from these two modes are combined by considering the difference of uncertainties for these two modes.
The combination of common and independent systematic uncertainties from the first category is calculated using the weighted least squares method~\cite{DAgostini:1993arp}.
The combined systematic uncertainty from the first category is 7.0\%.
\begin{table}[htbp!]
  \centering
  \caption{Summary of the first part of the systematic uncertainties for mode I and mode II. 
  The items with * are common uncertainties for these two decay modes.
  }
   \label{tab:1st_sys_err}
    \begin{tabular}{lcc}
    \hline \hline
    Source & mode I & mode II \\ \hline
    MDC tracking* & 4.0 & 4.0 \\
    PID* & 4.0	& 4.0 \\
    Photon detection* & 5.0 & 4.0 \\
    Kinematic fit & 1.9 & 0.8 \\
    $\etap$ resolution & 0.1 & 0.4 \\
    ${\cal B}(\etap\rightarrow\eta\pp)$* &  1.0 & 0.5 \\
    ${\cal B}(\etap\rightarrow\gam\pp)$ &  ... & 0.4 \\
    ${\cal B}(\eta\rightarrow\gam\gam)$* &  0.4 & 0.2 \\
    Number of $\jpsi$ events* & 0.4 & 0.4 \\
    Total &	 7.9	& 7.0 \\
    \hline \hline
    \end{tabular}
\end{table}
The second source comes from the PWA fit procedure, where the systematic uncertainties are applied to measurements of the branching fractions and resonance parameters.
These sources of systematic uncertainties are described as below.
\begin{itemize}
	\item[(i)] BW parameterization. 
	The uncertainty from the BW parameterization is estimated by the changes in the fit results caused by replacing the constant width $\Gamma_0$ of the BW for $f_0(2020)$, which is close to the $\bietap$ threshold, with the mass-dependent width as explained in Sec.~\ref{sub:pwa_result}.
	\item[(ii)] Background uncertainty. 
	To estimate the background uncertainty, alternative fits are performed with background events from different $\etap$ sideband regions and different normalization factors, and the differences of the results are assigned as the systematic uncertainty. 
	\item[(iii)] Uncertainties from additional resonances. 
	Uncertainties from possible additional resonances are estimated by adding the $\rho(1900)$ and the $h_1(1170)$, which are the two most significant additional resonances, into the baseline configuration individually. 
	The changes of the results caused by them are assigned as the systematic uncertainties.
\end{itemize}
For each alternative fit performed to estimate the systematic uncertainties from the PWA fit procedure, the changes of the results are taken as the one-sided systematic uncertainties.
For each measurement, the individual uncertainties are assumed to be independent and are added in quadrature to obtain the total systematic uncertainty on the negative and positive side, respectively.
The sources of systematic uncertainties are applied to the measurements of masses and widths of all resonances, and their contributions are summarized in Tab.~\ref{tab:un_sum_mw}.
\begin{table*}[htbp]
\small
\centering
\caption{Summary of the systematic uncertainties on the masses (MeV/$c^{2}$) and widths (MeV) of the resonances in the baseline model, denoted as $\Delta$M and $\Delta\Gamma$.}
\label{tab:un_sum_mw}
\setlength{\tabcolsep}{2.0 mm}{
	\renewcommand\arraystretch{1.5}
	\begin{tabular} { l c   c   c c c c  c c c c}
	\hline \hline 
	\multirow{2}*{Sources} & \multicolumn{2}{c }{$f_0(2020)$}  &  \multicolumn{2}{c }{$f_0(2330)$} & \multicolumn{2}{c }{$f_0(2480)$}  &  \multicolumn{2}{c}{$h_1(1415)$} & \multicolumn{2}{c}{$f_2(2340)$} \\
	\cline{2-3} \cline{4-5}  \cline{6-7} \cline{8-9} \cline{10-11}
	~ & $\Delta$M & $\Delta \Gamma $ & $\Delta$M & $\Delta \Gamma $ & $\Delta$M  & $\Delta \Gamma $ & $\Delta$M   & $\Delta \Gamma $ & $\Delta$M  & $\Delta \Gamma $ \\ 
		\hline
	Breit-Wigner formula &	+45 &	+26 &	+2 &	+29 & $-6$  & +11 & +1 & +3 & $-1$ & $-4$ \\ 	
	Extra resonances  &	+30 & ${}^{+35}_{-49}$ & +8	& $-9$ & +4 & $-8$ & +8 & $-10$ & ${}^{+22}_{-6}$ & ${}^{+26}_{-6}$\\ 	
	Background uncertainty  & +2	& +14	& +5	& +7	& +1 & +3 & +4 & +12 & +1 & $-10$ \\
	Total & +54 & ${}^{+46}_{-49}$ & $+10$ & ${}^{+30}_{-9}$ & ${}^{+4}_{-6}$ & ${}^{+11}_{-8}$ & $+9$ & ${}^{+12}_{-10}$ & ${}^{+22}_{-6}$ & ${}^{+26}_{-12}$ \\	
	\hline \hline
	\end{tabular} 
	}
\end{table*}
The relative systematic uncertainties relevant to the branching fraction measurements are summarized in Tab.~\ref{tab:un_sum_br}, where the last row lists the total relative systematic uncertainties from fitting irrelevant sources.
\begin{table*}[htbp!]
\centering
\caption{Summary of systematic uncertainties on the branching fraction of $\jpsi\rightarrow\gam~X,~X\rightarrow\etap\etap$ or $\jpsi\rightarrow\etap~X,~X\rightarrow\gam\etap$ (\%).}
\label{tab:un_sum_br}
\setlength{\tabcolsep}{2.0 mm}{
	\renewcommand\arraystretch{1.5}
	\begin{tabular} { l   c   c   c c c c }
	\hline \hline 
	Sources & $f_0(2020)$ & $f_0(2330)$ & $f_0(2480)$ & $h_1(1415)$ & $f_2(2340)$ & $0^{++}$ PHSP \\
	\hline
	First category &  \multicolumn{6}{c}{$\pm$7.0}	\\
	Breit-Wigner formula &	+6.7 & +63.4	& +40.7	& $-5.2$ & $-2.9$ & +38.0 	\\	
	Extra resonances  &	$-16.3$ & $-26.8$	& $-26.4$ & $-37.9$ & ${}^{+0.8}_{-17.5}$  & ${}^{+347.8}_{-59.1}$ \\ 		
	Background uncertainty  & +6.8	& +15.4 & +19.4 & +14.0 & $-2.9$ & +3.0 \\ 	
	Total & ${}^{+11.8}_{-17.7}$ & ${}^{+65.7}_{-27.7}$ & ${}^{+45.6}_{-27.3}$ & ${}^{+15.7}_{-38.9}$ & ${}^{+7.1}_{-19.3}$ & ${}^{+350.0}_{-59.5}$ \\
	\hline \hline
	\end{tabular} 
	}
\end{table*}

\section{Summary}
In summary, a PWA of the decay $\jpsi\rightarrow\gam\bietap$ has been performed based on $10.09\times10^9~\jpsi$ events collected with the BESIII detector.
The dominant contributions are from the scalars, $f_0(2020)$ and $f_0(2330)$, together with a
non-negligible contribution from $f_2(2340)$.
The $h_1(1415)$ in $\gam\etap$ is also observed.
Their measured masses and widths are consistent with the PDG values~\cite{Zyla:2020zbs}.
A possible new $0^{++}$ state, $f_0(2480)$, with a significance of 5.2$\sigma$, whose mass and width are respectively $2470\pm4^{+4}_{-6}$~MeV/$c^{2}$ and $75\pm 9 ^{+11}_{-8}$~MeV, is needed to describe the data.

The $f_0(2020),~f_0(2330)$, and $f_2(2340)$ are observed in the $\etap\etap$ decay mode for the first time.
The production of $f_0(2020)$ in $\jpsi\rightarrow\gam\etap\etap$ is compatible with that of $f_0(2100)$ in $\jpsi\rightarrow\gam\eta\eta$~\cite{Ablikim:2013hq} and $f_0(2200)$ in $\jpsi\rightarrow\gam\ksks$~\cite{Ablikim:2018izx}.
If those scalars are assigned to one resonance in this analysis, its large production rate in radiative $\jpsi$ decay 
suggests that it has a large overlap with scalar glueball.
However, its mass is lower than the mass of the first excitation of scalar glueball from the LQCD prediction~\cite{Bali:1993fb,Morningstar:1999rf,Chen:2005mg,Gregory:2012hu,Sun:2017ipk}.
With the high statistics $\jpsi$ data sample collected with BESIII, it is now critical to study the pole structure of the scalars around 2.1~GeV/$c^2$ with coupled channel analyses~\cite{Rodas:2021tyb,Sarantsev:2021ein}.

\begin{acknowledgments}
The BESIII collaboration thanks the staff of BEPCII and the IHEP computing center for their strong support. This work is supported in part by National Key Basic Research Program of China under Contract Nos. 2020YFA0406300, 2020YFA0406400; 
National Natural Science Foundation of China (NSFC) under Contracts Nos. 11625523, 11635010, 11675183, 11735014, 11822506, 11835012, 11922511, 11935015, 11935016, 11935018, 11961141012, 12022510, 12025502, 12035009, 12035013, 12061131003; 
the Chinese Academy of Sciences (CAS) Large-Scale Scientific Facility Program; 
Joint Large-Scale Scientific Facility Funds of the NSFC and CAS under Contracts Nos. U1732103, U1732263, U1832207; 
CAS Key Research Program of Frontier Sciences under Contracts Nos. QYZDJ-SSW-SLH003, QYZDJ-SSW-SLH040; 
100 Talents Program of CAS; 
INPAC and Shanghai Key Laboratory for Particle Physics and Cosmology; 
ERC under Contract No. 758462; 
European Union Horizon 2020 research and innovation programme under Contract No. Marie Sklodowska-Curie grant agreement No 894790; 
German Research Foundation DFG under Contracts Nos. 443159800, Collaborative Research Center CRC 1044, FOR 2359, FOR 2359, GRK 214; 
Istituto Nazionale di Fisica Nucleare, Italy; 
Ministry of Development of Turkey under Contract No. DPT2006K-120470; National Science and Technology fund; 
Olle Engkvist Foundation under Contract No. 200-0605; 
STFC (United Kingdom); The Knut and Alice Wallenberg Foundation (Sweden) under Contract No. 2016.0157; 
The Royal Society, UK under Contracts Nos. DH140054, DH160214; The Swedish Research Council; 
U. S. Department of Energy under Contracts Nos. DE-FG02-05ER41374, DE-SC-0012069.
\end{acknowledgments}

\bibliography{reference}

\end{document}